\documentclass{aa}  

\usepackage{graphicx}
\usepackage{txfonts}
\usepackage[]{hyperref}
\usepackage{xcolor}

\begin{document}

   \title{Newly identified compact hierarchical triple system candidates using \textit{Gaia} DR3}

   \author{D.~R.~Czavalinga
          \inst{1,2}\fnmsep\thanks{Email: czdonat@titan.physx.u-szeged.hu}
          \and
          T.~Mitnyan\inst{1,2}
          \and  
          S.~A.~Rappaport\inst{3}
          \and
          T.~Borkovits\inst{1,2,4,5,6}
          \and
          R.~Gagliano\inst{7}
          \and
          M.~Omohundro\inst{8}
          \and
          M.~H.~K.~Kristiansen\inst{9}
          \and
          A.~P\'al\inst{4}
          }

   \institute{Baja Astronomical Observatory of University of Szeged, H-6500 Baja, Szegedi út, Kt. 766, Hungary
   \and
   ELKH-SZTE Stellar Astrophysics Research Group, H-6500 Baja, Szegedi út, Kt.
766, Hungary
         \and
Department of Physics, Kavli Institute for Astrophysics and Space Research, M.I.T., Cambridge, MA 02139, USA
\and
Konkoly Observatory, Research Centre for Astronomy and Earth Sciences, H-1121 Budapest, Konkoly Thege Miklós út 15-17, Hungary
\and
ELTE Gothard Astrophysical Observatory, H-9700 Szombathely, Szent Imre h. u. 112, Hungary
\and
MTA-ELTE Exoplanet Research Group, H-9700 Szombathely, Szent Imre h. u. 112, Hungary
\and
Amateur Astronomer, Glendale, AZ 85308, USA
\and
Citizen Scientist, c/o Zooniverse, Department of Physics, University of Oxford, Denys Wilkinson Building, Keble Road, Oxford, OX1 3RH, UK
\and
Brorfelde Observatory, Observator Gyldenkernes Vej 7, DK-4340 T\o ll\o se, Denmark
}

   \date{Received ...; accepted ...}

 
  \abstract
   {}
   {We introduce a novel way to identify new compact hierarchical triple stars by exploiting the huge potential of \textit{Gaia} DR3 and also its future data releases. We aim to increase the current number of compact hierarchical triples significantly.}
   {We utilize several eclipsing binary catalogs from different sky surveys totaling more than 1 million targets for which we search for \textit{Gaia} DR3 Non-single Star orbital solutions with periods substantially longer than the eclipsing periods of the binaries. Those solutions in most cases should belong to outer orbits of tertiary stars in those systems. We also try to validate some of our best-suited candidates using \textit{TESS} eclipse timing variations.}
   {We find 403 objects with suitable \textit{Gaia} orbital solutions of which 27 are already known triple systems. This makes 376 newly identified hierarchical triple system candidates in our sample. We analyze the cumulative probability distribution of the outer orbit eccentricities and find that it is very similar to the ones found by earlier studies based on the observations of the \textit{Kepler} and OGLE missions. We found measurable non-linear eclipse timing variations or third-body eclipses in the \textit{TESS} data for 192 objects which we also consider to be confirmed candidates. Out of these, we construct analytical light-travel time effect models for the eclipse timing variations of 22 objects with well-sampled \textit{TESS} observations. We compare the outer orbital parameters from our solutions with the ones from the \textit{Gaia} solutions and find that the most reliable orbital parameter is the orbital period, while the values of the other parameters should be used with caution.}
   {}


   \maketitle
%

\section{Introduction}
\label{introduction}
Hierarchical triple stellar systems consist of three stars that are organized as an inner, binary pair with a more distant, outer (tertiary) component. These systems are exceptional in that their orbital and astrophysical parameters can often be determined precisely, especially for those in which the inner binary either eclipses, or is eclipsed by, the tertiary during their orbital revolution. These highly accurate parameters can be used to refine stellar evolutionary models and to study the formation and evolution of such systems from which all fields of stellar astrophysics can benefit. These systems can also be of help in the understanding of some peculiar objects with compact components \citep[e.g. Type Ia supernovae,][]{kushnir13} because the formation of their progenitor close binaries requires some effective orbit shrinking mechanism during their lifetime that in many cases could be explained by dynamical interactions with a third stellar component in the system. Recent reviews about the formation and evolutionary scenarios of triple systems can be found in \citet{tokovinin21}, and \citet{toonen20,toonen22}, respectively.

These so-called dynamical interactions are more common and easier to detect within the scale of human lifetimes in compact hierarchical triple (CHT) systems where the outer orbital period is less than $\sim$1000 days. There is still a relatively small number of systems known in this category; the latest version of the Multiple Star Catalog \citep{tokovinin18} lists 201 such systems, but there are also a few quadruple or higher order hierarchical systems included in this number as well. Nevertheless, nowadays, thanks to the various ground-based and more importantly space-based photometric all-sky surveys, the numbers of compact hierarchical systems are rapidly growing. For example, \citet{hajdu19} and \citet{hajdu22} identified 177 and 16 such CHTs, respectively, utilizing the data of the OGLE (Optical Gravitational Lensing Experiment) survey. A recent comprehensive review of the different methods and results of the field of discovering compact hierarchical systems can be found in \citet{borkovits22}.

Most of the CHT stellar systems were discovered by either photometric or spectroscopic studies and the detection of the astrometric wobble of any components of such systems was not feasible as part of broad searches. The third data release \citep{gaia22b,babusiaux22} of the \textit{Gaia} space telescope \citep{gaia16} is unique in the sense that it contains a catalog of objects for which the time-domain, ultra-precise astrometric observations allowed the \textit{Gaia} team to constrain a simple Keplerian two-body orbital solution for 135\,760 objects utilizing their variable astrometric positions on the plane of the sky. Moreover, this data release also contains the first spectroscopic results of \textit{Gaia} in which the time-domain radial velocity measurements also resulted in 186\,905  objects with a similar kind of two-body orbital solutions. Finally, there are some objects in which eclipses could be detected in the \textit{Gaia} photometric measurements and these eclipses made it possible to construct such a two-body orbital solution for 86\,918 objects. The previously mentioned solutions are of a single type, but there are also systems with two types of observational data for which a combined orbital solution was found, namely `astrometric+spectroscopic' (33\,467 objects), `eclipsing+spectroscopic' (155 objects). Detailed information about the modeling steps can be found in the corresponding section of the \textit{Gaia} DR3 Online documentation \citep{pourbaix22}. All of the orbital solutions and associated parameters for these binary or multiple star candidates are publicly available and tabulated in the Vizier tables of the \textit{Gaia} DR3 Non-single stars (NSS) catalog \citep{gaia22a}.

Our goal was to identify possible NSS models in which the orbital solution may belong to an outer orbit of a potential hierarchical triple (or multiple) system. These are detected through either the motion of the unresolved inner binary, the outer tertiary star, or the light centroid of both. For this purpose, our main idea was to collect as many eclipsing binaries with previously determined orbital periods from different catalogs found in the literature and then to search for \textit{Gaia} orbital solution periods that are at least a factor of five longer than the corresponding EB period. In Sect.~\ref{analysis}., we describe the data sets we used for our search along with the steps of their analysis. In Sect.~\ref{results}., we summarize and discuss the main results of our search for triple candidates, while in Sect.~\ref{validation}., we present a possible validation method of the newly found candidates using eclipse timing variations (ETVs) calculated from \textit{TESS} data. Finally, we highlight our main conclusions in Sect.~\ref{conclusions}.

\section{Catalog data and analysis -- Cross-matching and identification of candidates}
\label{analysis}

We downloaded the full data sets of all main catalogs containing lists of eclipsing binaries (EBs) namely APASS \citep[AAVSO Photometric All Sky Survey; ][]{munari2014}, ASAS-SN \citep[All-Sky Automated Survey for Supernovae; ][]{shappee14,rowan22}, GCVS \citep[General Catalog of Variable Stars; ][]{samus17}, \textit{Kepler} \citep[\textit{Kepler} Eclipsing Binary Stars; ][]{kirk16}, \textit{TESS} \citep[\textit{TESS} Eclipsing Binary Stars; ][]{prsa22} and VSX \citep[The International Variable Star Index; ][]{watson06}. The total number of objects exceeded a million EBs (see Table~\ref{catalogs}). We cross-matched these lists with the \textit{Gaia} EDR3 source catalog\footnote{The \textit{Gaia} EDR3 source IDs of objects remained the same in the \textit{Gaia} DR3 catalog.} by coordinates via TOPCAT \citep[Tool for OPerations on Catalogues And Tables;  ][]{taylor05}.  We used the default 5\,\arcsec search radius for this purpose, which yielded more sources than the original lists implying a number of duplicates. We did not filter these out right away as it was easier and more efficient to remove them at the end of the analysis (see the text later).

\begin{table}
\caption{Number of EBs in different catalogs and in total for which we carried out our analysis including multiple occurrences (see the text for details). }             
\label{catalogs}      
\centering                          
\begin{tabular}{c | c}        
\hline\hline                 
Catalog & Number of EBs \\    
\hline                        
   APASS & 4\,516 \\      
   ASAS-SN & 35\,464 \\
   GCVS & 6\,849 \\
   \textit{Kepler} EBs & 2\,876 \\
   \textit{TESS} EBs & 4\,584 \\ 
   VSX & 971\,757 \\
\hline                                   
\hline
   Total & 1\,026\,046 \\
\end{tabular}
\end{table}

After cross-matching with the \textit{Gaia} EDR3 catalog, we downloaded all the Vizier tables in the NSS catalog containing the different types of NSS solutions. We then performed a search for each \textit{Gaia} source ID that was in our lists and collected all available orbital model parameters for the source. In order to make a more complete survey, we also searched for stars with a relatively small angular separation\footnote{A hundred times the angular diameter of the orbit of the corresponding \textit{Gaia} NSS solution.} from the unresolved EBs in our list to see if we could find a bound tertiary component on a wider orbit around them that is resolved by \textit{Gaia}. However, we could not find any system that met such criteria, mainly because of the primary selection of the astrometric processing steps applied by the \textit{Gaia} team that makes it unlikely to find any such systems \citep{halbwachs22,riello21}.

As a next step, we kept all objects for which the \textit{Gaia} orbital period solution was at least five times longer than the orbital period found in the corresponding EB catalog. We chose this condition as a very loose threshold because currently the tightest known triple system has an outer-to-inner period ratio of 5.4 \citep{xia19} and because of dynamical stability issues. It is very unlikely that we will find one below our chosen period ratio limit \citep[see][]{borkovits22a}. Finally, in order to filter out duplicate sources, we joined the list of objects we kept from the different catalogs and filtered out those source IDs that appeared in this complete list multiple times (so those that are listed in multiple catalogs under different names) to have only a single occurrence of each distinct object. We also got rid of those systems for which multiple different \textit{Gaia} source IDs had been found and kept only those for which the brightness were consistent with the brightness of the EB we searched for. We also noticed that some EBs had been listed with incorrect coordinates since we found different \textit{Gaia} source IDs for them based on those incorrect coordinates than their real source IDs. We found their real source IDs based on their common names instead of their coordinates listed in the corresponding catalogs. We threw out these systems as well, because neither of them actually had a \textit{Gaia} NSS solution that fulfilled our search criteria.

\section{Results}
\label{results}

\begin{figure*}
\caption{Distribution of inner binary periods and the corresponding outer orbital periods for our newly identified hierarchical triple star candidates. The dashed line shows the limiting period ratio of 5 chosen by us based on the requirement for dynamical stability. The additional panels on top and on the right show the one-dimensional distributions of the
periods. Different colors represent different types of \textit{Gaia} NSS orbital solutions.}             
\label{period_distribution}
    \centering
    \includegraphics[scale = 1.00]{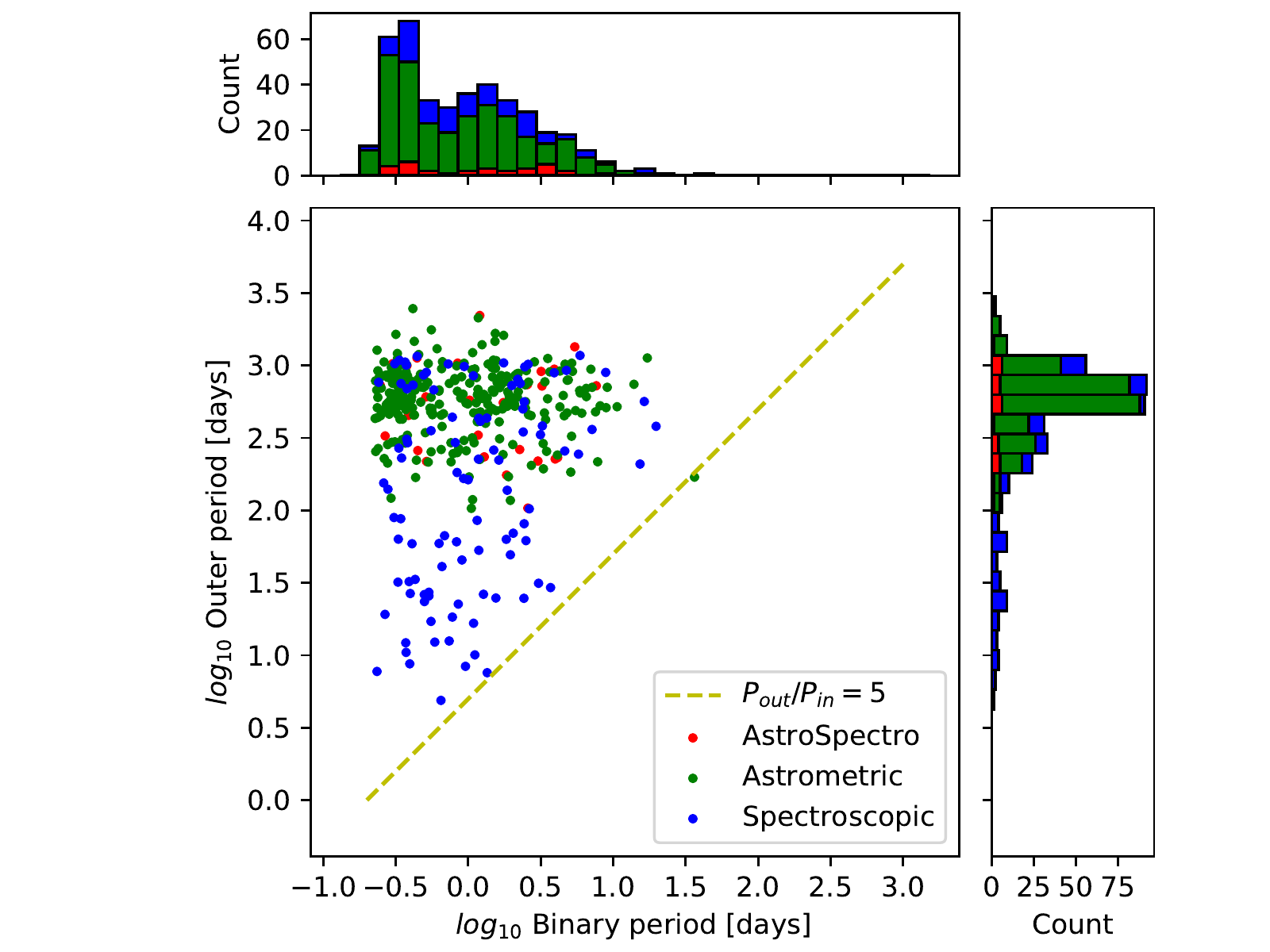}
\end{figure*}

\begin{figure}
\caption{{\it Top panel:} Differential probability distribution of the outer orbit eccentricities for our newly identified hierarchical triple star candidates. Different colors represent different types of \textit{Gaia} NSS
orbital solutions. The magenta line shows the analytical function described in Eq.~\ref{eq1}. {\it Bottom panel:} Cumulative eccentricity distributions.  Green - OGLE; blue - {\it Kepler}; red - {\it Gaia} (this study); purple - $dN/de =$ constant; black - a so-called `thermal' distribution with $dN/de \propto e$.}           
\label{ecc_distribution}
    \centering
    \includegraphics[width = 0.5 \textwidth]{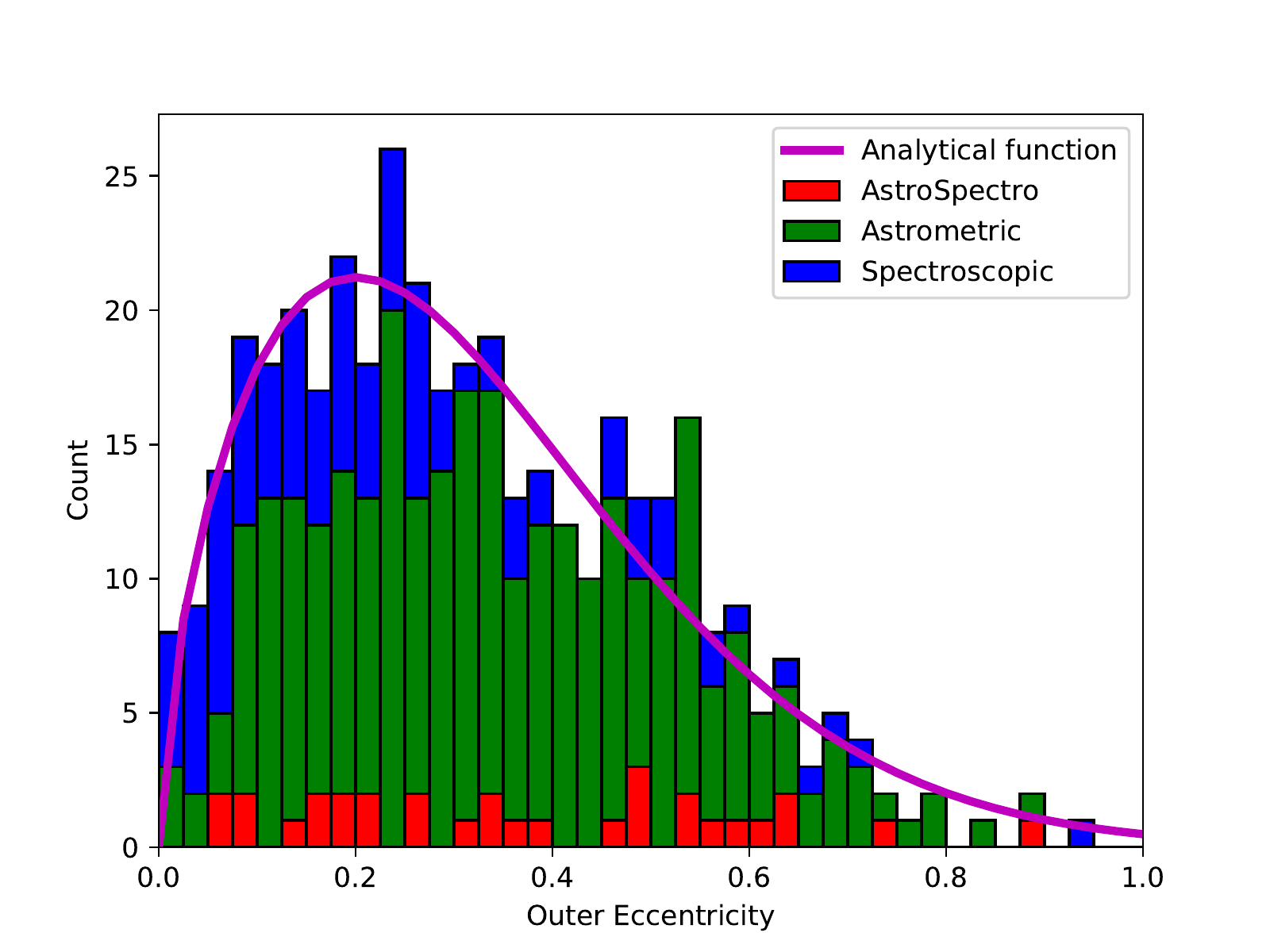}
    \includegraphics[width = 0.45 \textwidth]{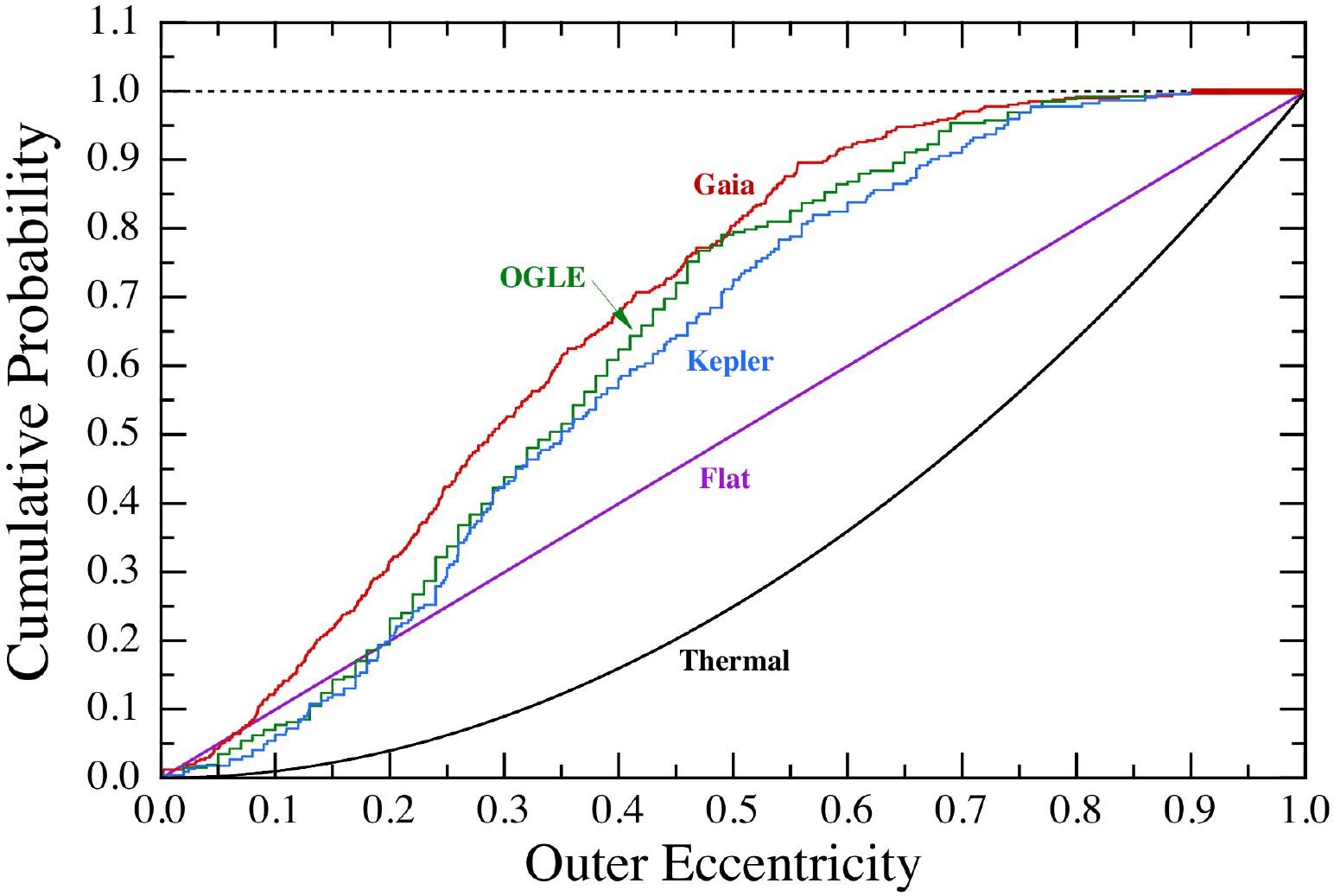}
\end{figure}
\subsection{Summary of the identified candidates}
\label{summary}
As a result of our study, we have identified 403 hierarchical triple candidates among formerly known EBs from the literature based on the periods of their \textit{Gaia} NSS orbital solutions. The first 35 candidates and the parameters of their corresponding \textit{Gaia} NSS orbital solutions are listed in Table~\ref{list_of_candidates}, while the full list will be available as an online table in CDS. There are 100 pure spectroscopic, 267 pure astrometric and 31 combined \textit{Gaia} NSS solutions in our sample. There are also three objects with two separate astrometric and spectroscopic solutions along with two other objects with separate astrometric and combined solutions. For objects with two different types of orbital solutions, we chose to use the pure astrometric solution in every case as a convention. In Fig.~\ref{period_distribution} we illustrate the distribution of the inner and outer orbital periods (the periods of the corresponding NSS solutions) for all of our candidates. Broadly speaking, the outer periods range from somewhat greater than 1000 days, down to just a few days, though with the bulk of them concentrated between 300 and 1000 days.  The sharp falloff in systems above $\sim$1000 days is an artefact of the \textit{Gaia} observational window. The inner binary periods range from a fraction of a day to 10 days, simply representing the observational selection effects of the underlying EB catalogs that we utilized.  Most of the longer outer periods are based on either a \textit{Gaia} astrometric or spectroscopic solution, while essentially all of the outer orbits with periods less than $\sim$100 days are pure \textit{Gaia} spectroscopic solutions.  We also indicate our chosen limiting period ratio of $P_{\rm out}/P_{\rm in} = 5$ with a dashed line.  The closer a system is to this line, the tighter it is. We note that one point is below this limit for which we will give an explanation later in Sect.~\ref{validated_sufficient}.

Prior to this study, there were some 400 known triple-star systems with outer periods $\lesssim 1000$ days, and only about 30 known with periods $\lesssim 200$ days.  These systems are discussed and tabulated in \citet{tokovinin18,borkovits16,hajdu19,hajdu22} and Fig.~1 of \citet{borkovits22a}.  Thus, this study has the potential to  substantially increase the number of known triple-star systems with measured outer orbits.

In Fig.~\ref{ecc_distribution} (upper panel) we show the differential eccentricity probability distribution for the 403 triple candidates that we have identified.  The distribution is crudely described by the function
\begin{equation}
\label{eq1}
dN/de \simeq 8.34 \,e^{0.618} \exp[-5.15 e^{1.618}]
\end{equation}
where $e$ is the outer eccentricity (see below for the origin of this functional form).  This curve is superposed (after renormalization with the number of objects and the bin width) on the eccentricity distribution in Fig.~\ref{ecc_distribution}, and provides a decent fit, considering the limited statistics. In the lower panel of Fig.~\ref{ecc_distribution} we show the cumulative probability distribution for the outer eccentricities that we have found.  For comparison, we also show the eccentricity distributions for the {\it Kepler} sample of triples \citep{borkovits16} and the outer orbits for 205 secure triples identified in the OGLE sample \citep{hajdu19}.  In the same plot we compare these three empirical distributions with those of a flat eccentricity distribution ($dN/de =$~constant) and a so-called `thermal distribution' ($dN/de \propto e$).  It seems apparent that the empirical distributions are fairly similar to each other, but distinctly different from a differential distribution that is constant (linear as a cumulative distribution).  And, for what it is worth, the empirical distributions are dramatically distinct from a thermal distribution, as suggested by \citet{jeans19} for binary stellar systems which have reached a state of energy equipartition through a large series of dynamical encounters.  In fact, the empirical distributions fairly closely resemble the empirical eccentricity distribution for ordinary binary stellar systems \citep[see Fig.~3 of][]{duchene13}.  

The cumulative eccentricity distribution for the \textit{Gaia}-based outer eccentricity distribution can be reasonably well represented by the following analytic expression:
\begin{equation}
\label{eq2}
N(>e) \simeq 1-\exp[-(e/0.363)^{1.618}].
\end{equation}
The derivative of this expression yields the above analytic approximation to the differential eccentricity distribution.  There is one observational selection effect that we can envision for this data set in regard to the eccentricity distribution. That is, large eccentricities generally tend to make the triple star system more unstable.  From Equation (16) of \citet{rappaport13} \citep[based on expressions for dynamical stability by][]{mikkola08,mardling01},
\begin{equation}
\label{eq3}
P_{\rm out}/P_{\rm in} \gtrsim 4.7 (1+e)^{3/5} (1-e)^{-9/5}
\end{equation}
we see that for period ratios of 1000, 300, 100, 30, and 10, the limiting eccentricities are $e = 0.94; 0.88; 0.78; 0.58$ and 0.28.  Thus, one has to be somewhat careful about interpreting the relative dearth of high eccentricity systems.

In the sections below we discuss what we have done to try to validate the triple-star nature of some of these systems using ETV data.  Additionally, as it turns out, some 27 of these systems were previously known triples or multiples (see Table~\ref{known_candidates}).  We also discuss the `truth probability' for the SB1 solutions evaluated from the empirical formula of \citep{bashi22}.  Thus, through these we acquire reasonable confidence that a substantial fraction (e.g., $\gtrsim 50\%$, and probably as high as 75\%) of the new candidates are actually valid triple stars with a small fraction of quadruples.  If this indeed turns out to be the case, then we will have substantially increased the number of known compact triple-star systems.   If even half of our new candidates are proven to be valid triples, then this study will have increased the number of known compact triples by $\sim$1.5 times both below 1000 days and below 200 days. These compact triple systems, with RV orbits, third-body eclipses, and/or measurable dynamical interactions are the most likely multi-stellar systems to yield comprehensive measurements of all their stellar and orbital parameters.  Here we hope to demonstrate by this project that finding \textit{Gaia} outer orbits associated with known eclipsing binaries is a novel way to find new compact triples.

\subsection{Systems already known as candidates}
\label{literature}
We checked the available publications in the literature for all single objects individually using SIMBAD \citep{wenger00} and NASA ADS\footnote{\href{https://ui.adsabs.harvard.edu/}{https://ui.adsabs.harvard.edu/}}. Most of these objects had not been involved in extensive studies apart from their identification as EBs. In Table~\ref{known_candidates}, we have listed all 27 systems among our candidates for which we found any information about their possible triple star nature in the literature. Comparing the outer orbital periods from the \textit{Gaia} solutions and the literature references, there are 14 objects (KIC 8330092, KIC 6525196, KIC 8043961, KIC 9665086, KIC 8719897, KIC 10991989, HD 181068, KIC 7955301, TIC 229785001, TIC 437230910, TIC 99013269, SAO 167450, HD 81919 and HD 190583) for which a reasonable agreement can be found. We consider these systems as validated candidates that are confirmed by the \textit{Gaia} mission as an independent source. The only exception is SAO 167450, that is a visual member of AA Ceti in a bound EB+EB (2+2) hierarchical quadruple system, for which the \textit{Gaia} spectroscopic orbital solution found belongs to one of the inner EBs and not to the outer orbit of the two EBs around their common center of mass.

For seven systems, although clear signs indicate that they have a multiple nature, their outer orbital period was unknown from previous studies. For V0857 Her, the presence of an additional component was suggested based on its rotational broadening function extracted from its spectra. For TIC 229082842 and TIC 59042291, an obvious sinusoidal variation was detected in their ETVs but the corresponding data sets were insufficient to derive their outer orbital periods. For TIC 305635022, signals of two eclipsing binaries were detected in its \textit{TESS} LC that originate at the same pixels, hence it is a candidate 2+2 system. The remaining three objects, TIC 24704862, TIC 301956407 and TIC 410654298 are all subsystems of known visual double or multiple stars. In our opinion, the orbital solutions found by \textit{Gaia} could also be considered as an independent confirmation of the multiple nature of these systems.

The rest of the six systems have significant differences between the two listed periods for them (\textit{Gaia} vs. the literature). Two of them, although cataloged as EBs, are actually single stars with transiting exoplanets (TIC 438071843 and HATS-26). For these systems, it is possible that the \textit{Gaia} orbital solutions belong to further stellar or exoplanet components on wider orbits. Finally, for the remaining four candidates, the ratios of the two periods (\textit{Gaia} vs. the literature) are close to integers or half-integers that could indicate that at least one of the solutions over/underfitted the available sparse data for the same orbits, or it is also possible that the listed orbital periods belong to different components in these systems.

\begin{table*}
\caption{Already known systems from the literature and their reference periods compared to the values from the corresponding \textit{Gaia} solutions (A: astrometric solution, S: spectroscopic solution, AS: combined solution).}             
\label{known_candidates}      
\centering                          
\begin{tabular}{c | c | c | c | c | c }        
\hline\hline                 
Common name &	\textit{Gaia} DR3 source ID & \begin{tabular}{@{}c@{}} \textit{Gaia} period \\ $\mathrm{[days]}$ \end{tabular} &	\begin{tabular}{@{}c@{}} Ref. period \\ $\mathrm{[days]}$ \end{tabular} &	Ref. &	Note
 \\    
\hline                        
GN Boo &	1281813580534856448 &	637.46 (A) &	3492.96 &	1, 2 & \\
V0899 Her &	1325067886936436224 & 454.30 (A) &	1351.43 &	3, 4 &	component A, maybe unresolved binary \\
V0857 Her &	1352346392464295168 & 292.87 (S) &	-- &	5 &	suggested companion based on BFs \\
TIC 99013269 &	1861222084858776576 &	609.52 (A) &	604.23 &	6 &	triply eclipsing triple \\
LO And &	1938276860462797568 &	329.92 (A) &	10782.18 &	7, 8 & \\
KIC 8330092 &	2076066725646925824 &	549.93 (A) &	581.00 &	9, 11 & \\
KIC 6525196 &	2077667962475652864 &	422.56 (A) &	418.20 &	10, 11 & \\
KIC 8043961 &	2078813069478750720 &	487.90 (A) &	478.60 &	10, 11 & \\
KIC 9665086 &	2080282635485293696 &	856.41 (A) &	856.00 &	9, 11 & \\
KIC 8719897 &	2082108645117526528 & 332.89 (S)	& 333.10 &	10, 11 & \\
KIC 10991989 &	2086499201207107072 &	550.64 (AS) &	548.00 &	10, 11 & \\
HD 181068 &	2101454419072479488 &	45.47 (S) &	45.47 &	12, 13 &	triply eclipsing triple \\
KIC 7955301 &	2126832448810860160 &	208.80 (S) &	209.10 &	10, 11 & \\
TIC 305635022 &	2195548425150787328 &	843.88 (A) &	-- &	14 &	double EB signal detected \\
TIC 229785001 &	2256269710706835712 &	165.99 (S) &	165.29 &	6 &	triply eclipsing triple \\
EM Psc &	2589993789205323136 &	844.67 (A) &	1205.33 &	15 & \\
TIC 229082842 &	3230233289730648576 &	847.00 (A) &	-- &	16 &	sinusoidal variation in eclipse timings \\
TIC 438071843 & 3368700905049734784 & 910.91 (A) &	2.62 &	17 &	exoplanet with 2.6d period \\
TIC 437230910 &	3549833939509628672 &	292.98 (A) &	293.94 &	18, 19 & \\
TIC 59042291 &	40119431248460032 &	1275.11 (A) & -- &	16 &	sinusoidal variation in eclipse timings \\
TIC 24704862 & 4697659516960103168 & 1126.71 (A) & -- & 20, 21 & subsystem of HD 10607\\
SAO 167450 &	5134682575449622656 & 25.68 (S) &	25.68 &	22, 23 &	EB period of the visual pair of AA Ceti \\
TIC 410654298 & 5236430419447179776 & 169.60 (AS) & -- & 24, 25 & subsystem of HD 102579\\
HD 81919 &	5437474574366697088 &	690.31 (S)	& 721.33 &	26, 27 &	component Aab \\
HATS-26 &	5656896924435896832 & 193.54 (A) &	3.30 &	28 &	hot Jupiter with 3 days period \\
TIC 301956407 & 5811053234948685312 & 63.15 (S) & -- & 29, 21 & subsystem of HD 155875\\
HD 190583 &	6374231714992810752 & 671.51 (A) &	662.50 &	30 & \\
\hline                                   
\end{tabular}
\tablebib{
(1) \citet{yang13}; (2) \citet{wang15}; (3) \citet{lu01}; (4) \citet{qian06}; (5) \citet{pribulla09}; (6) Visual Survey Group \citep[VSG;][]{vsg22}, unpublished; (7) \citet{gurol05}; (8) \citet{huang21}; (9) \citet{conroy14}; (10) \citet{rappaport13}; (11) \citet{borkovits16}; (12) \citet{derekas11}; (13) \citet{borkovits13}; (14) \citet{zasche22}; (15) \citet{qian08}; (16) \citet{lohr15}; (17) \citet{vanderburg16}; (18) \citet{pourbaix04}; (19) \citet{griffin06}; (20) \citet{luyten55}; (21) \citet{tokovinin14}; (22) Herschel (1782) according to the Washington Double Star Catalog \citep[WDS][]{mason01}; (23) \citet{fekel09}; (24) Gilliss (1868) according to the WDS \citep{mason01}; (25) \citet{tokovinin97}; (26) \citet{vandenbos21}; (27) \citet{tokovinin19}; (28) \citet{espinoza16}; (29) \citet{innes95}; (30) \citet{triaud17}
}
\end{table*}

\section{Validation of some systems using \textit{TESS} ETVs}
\label{validation}

In order to validate our list of new triple candidates with independent data and another method, we also collected \textit{TESS} Full-Frame Images (FFIs) for all objects from every available Sector up through 55. The light curves (LCs) were extracted from the FFIs in an automated manner applying a convolution-aided image subtraction photometry pipeline \citep[see e.g. ][]{mitnyan20} using FITSH \citep{pal12}. The remaining non-astrophysical trends were mostly removed from the LCs by the WOTAN package \citep{hippke19}. In the case of LCs that were affected with significant straylight or an extra (probably false) signal from a nearby object, we performed a Principal Component Analysis (PCA) that is more powerful in removing signals that originate outside the aperture used for the photometry. For this purpose, we used the built-in methods of the lightkurve \citep{lightkurve18} Python package and its corresponding dependencies: astropy \citep{astropy18}, astroquery \citep{ginsburg19} and TESScut \citep{brasseur19}.  After that, we calculated the ETVs for each object from all their available eclipses found in the \textit{TESS} LCs in the same way as described in \citep{borkovits15}. During this process, we found that some of the EBs are cataloged with incorrect eclipsing periods, so we listed those systems in Table~\ref{revised_periods} along with their revised EB periods based on \textit{TESS} LCs.

Next, we calculated a simple analytic light-travel time effect (LTTE) model for all the available ETVs using the orbital parameters of the corresponding \textit{Gaia} NSS solutions with the following equation: 

\begin{equation}
\label{eq4}
    \Delta_{\mathrm{LTTE}}=-\frac{a_{\mathrm{AB}}\sin{i_2}}{c} \frac{(1-e_{2}^{2})\sin{(\nu_2+\omega_2)}}{1+e_2\cos{\nu_2}},
\end{equation}
where $\Delta_{\mathrm{LTTE}}$ is the contribution of the LTTE to the ETVs, $a_{\mathrm{AB}}$ is the semi-major axis of the orbit of the inner binary component around the center of mass of the system, $i_2$ is the inclination, $e_2$ is the eccentricity, $\omega_2$ is the argument of periastron of the tertiary star in its outer orbit, $\nu_2$ is the true anomaly of the orbit of the tertiary star and $c$ is the velocity of light. We note that we also estimated the expected contributions of the so-called dynamical delays with Eq.~12 of \citet{borkovits16} which were substantially smaller than the amplitude of the LTTE, hence a simple LTTE model should be able to fit the ETVs well.

After that, we compared these LTTE models pre-calculated with the parameters from the \textit{Gaia} solutions to the ETVs from \textit{TESS} LCs. As our epoch and period determination are not always fully accurate, we also fitted a simple linear term for every model. As a first look, we noticed that none of the ETVs can be completely modeled with the parameters from the \textit{Gaia} orbital solutions without slight changes. Nevertheless, for the majority of the systems the outer orbital period of the \textit{Gaia} solution seemed to be in a good agreement with the ETVs.

\begin{figure*}
\caption{\textit{TESS} ETVs and their corresponding LTTE models for two of our validated triple candidates. Delay times calculated from primary and secondary eclipses are indicated with red and blue points, respectively, while their averaged values for the same orbital cycles are shown with black points. The green dashed lines denote the LTTE models calculated with the orbital parameters from the \textit{Gaia} NSS solution, while the blue lines represent our LTTE models after modifications in the orbital parameters. $P_0$ and $T_0$ are the eclipsing period and the epoch values, respectively, used to calculate the ETVs.\vspace{0.2cm}}             
\label{validated_etvs_1}
    \centering
    \includegraphics[scale = 0.6]{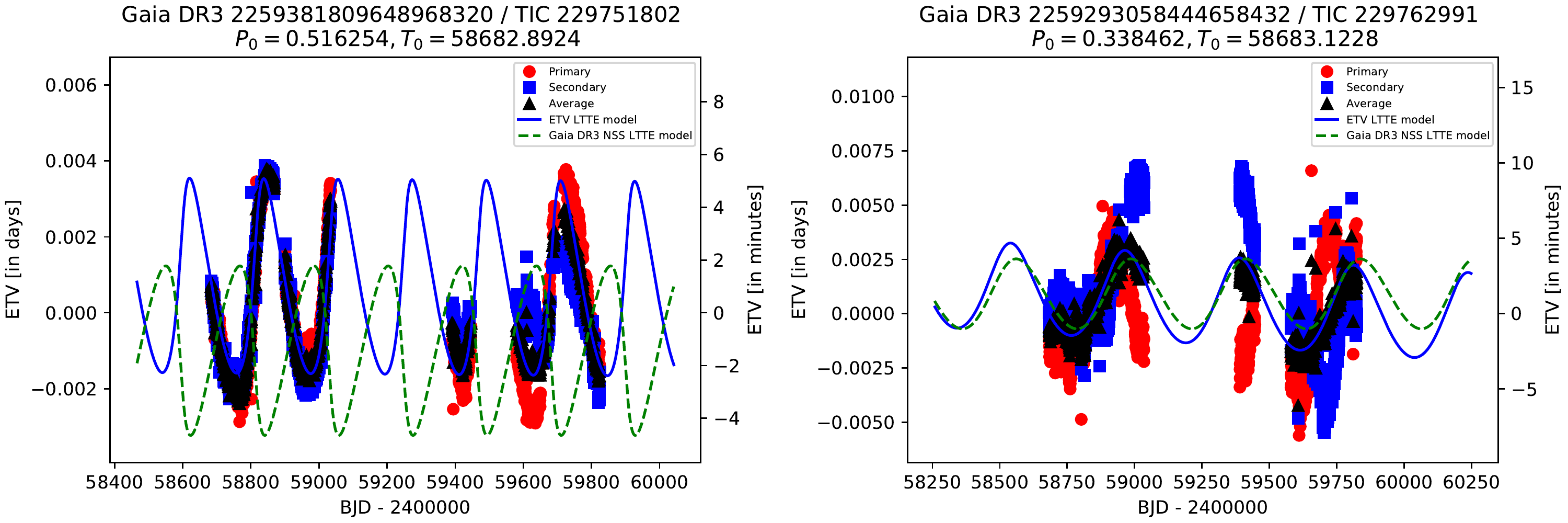}
\end{figure*}

As a next step, we tried to apply modifications to the parameters of the \textit{Gaia} NSS solution parameters in order to match the model LTTE curves to the observed \textit{TESS} ETVs. We tried to fit for the orbital parameters using a simple least squares method, but because of the very limited temporal coverage of the ETV points available for most objects, the majority of these fits resulted in significantly different, but equally acceptable solutions which heavily depended on the initial model parameters. Hence, we tried a different approach to fit the ETVs manually. For this purpose, we developed a Python Graphical User Interface (GUI) utilizing the PyQt5\footnote{\href{https://pypi.org/project/PyQt5/}{https://pypi.org/project/PyQt5/}} module that allows us to change the model parameters individually and plot the resulting models interactively in real-time, in order to find by visual inspection the best candidates with sufficient orbital coverage to perform a fit. Then, we tweaked the values of the parameters from the \textit{Gaia} solutions manually via a `chi-by-eye', in order to approximately match the ETV points, then optimized the parameters further with the Nelder-Mead Simplex method using the scipy\footnote{\href{https://docs.scipy.org/doc/scipy/index.html}{https://docs.scipy.org/doc/scipy/index.html}} Python package. Finally, we used these optimized parameters as initial parameters for a Markov-chain Monte Carlo (MCMC) sampling utilizing the emcee \citep{foreman-mackey13} Python package. 

\subsection{Validated candidates with sufficient orbital coverage or third-body eclipses}
\label{validated_sufficient}

The observing strategy of \textit{TESS} strongly limits the available data for objects, as most of them are observed only once or twice over $\sim$27 day long time-windows within a \textit{TESS} Year, but objects closer to the ecliptic poles can have more overlapping Sectors with better data coverage with up to a year long quasi-continuous window. Moreover, the same hemispheres are only observed every other year, hence an apparent one-year long gap will appear even in the data set of the most well-observed objects. This means that our candidates also suffer from these observational effects and as a result we could not confirm the triple nature of all of them even if they show significant non-linear ETVs.

Nevertheless, we found 22 candidates that have sufficient data coverage, so that their ETVs can be reliably modeled with a simple LTTE solution. We plot the ETVs and the corresponding simple LTTE solutions using the \textit{Gaia} and our own adjusted parameters for two of them in Fig.~\ref{validated_etvs_1} as examples, while the rest of them can be found in Figs.~\ref{validated_etvs_2},~\ref{validated_etvs_3} and \ref{validated_etvs_4} in the Appendix. The corresponding model parameters can be found in Table~\ref{LTTE_parameters}.

\begin{figure*}
\caption{Correlations between the orbital periods (upper left panel), projected semi-major axes (upper right panel), eccentricities (lower left panel) and argument of periastrons (lower right panel) coming from the \textit{Gaia} NSS solutions and their modified values after fitting LTTE models to the \textit{TESS} ETVs of our 22 validated triple candidates. Different colors signify different types of \textit{Gaia} NSS orbital solutions. The yellow dashed lines represent a ratio of unity, while for the bottom right panel the green dashed line shows the $\pm180\degr$ difference between the argument of periastrons coming from the two kinds of solutions. We note that in order to avoid breaking the dashed lines in the bottom right panel, we applied a $\pm360\degr$ shift for some of the $\omega_{\mathrm{out,NSS}}$ values which is equivalent because of the periodicity of the argument of periastron.}             
\label{parameter_correlations}
    \includegraphics[scale = 0.6]{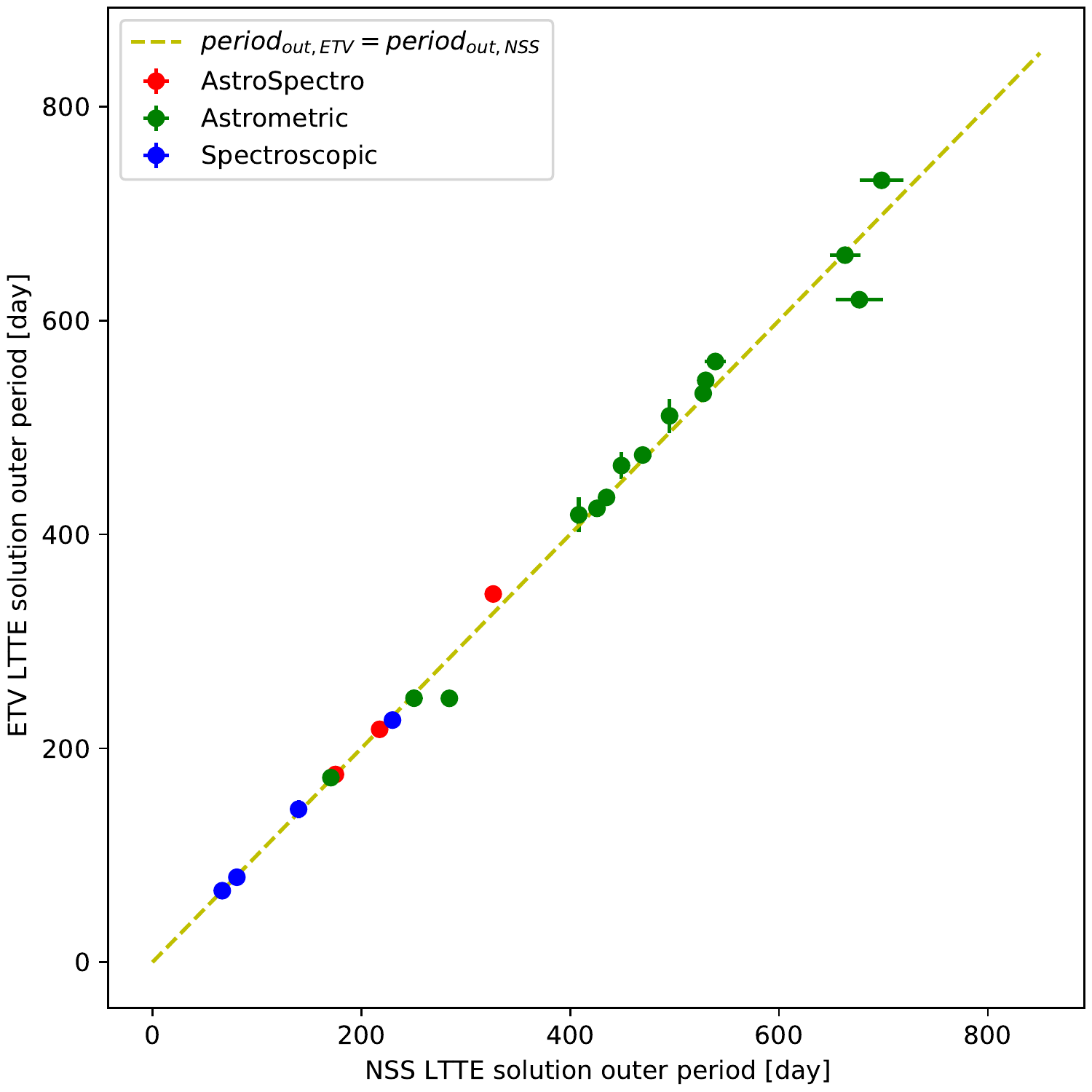}
    \includegraphics[scale = 0.6]{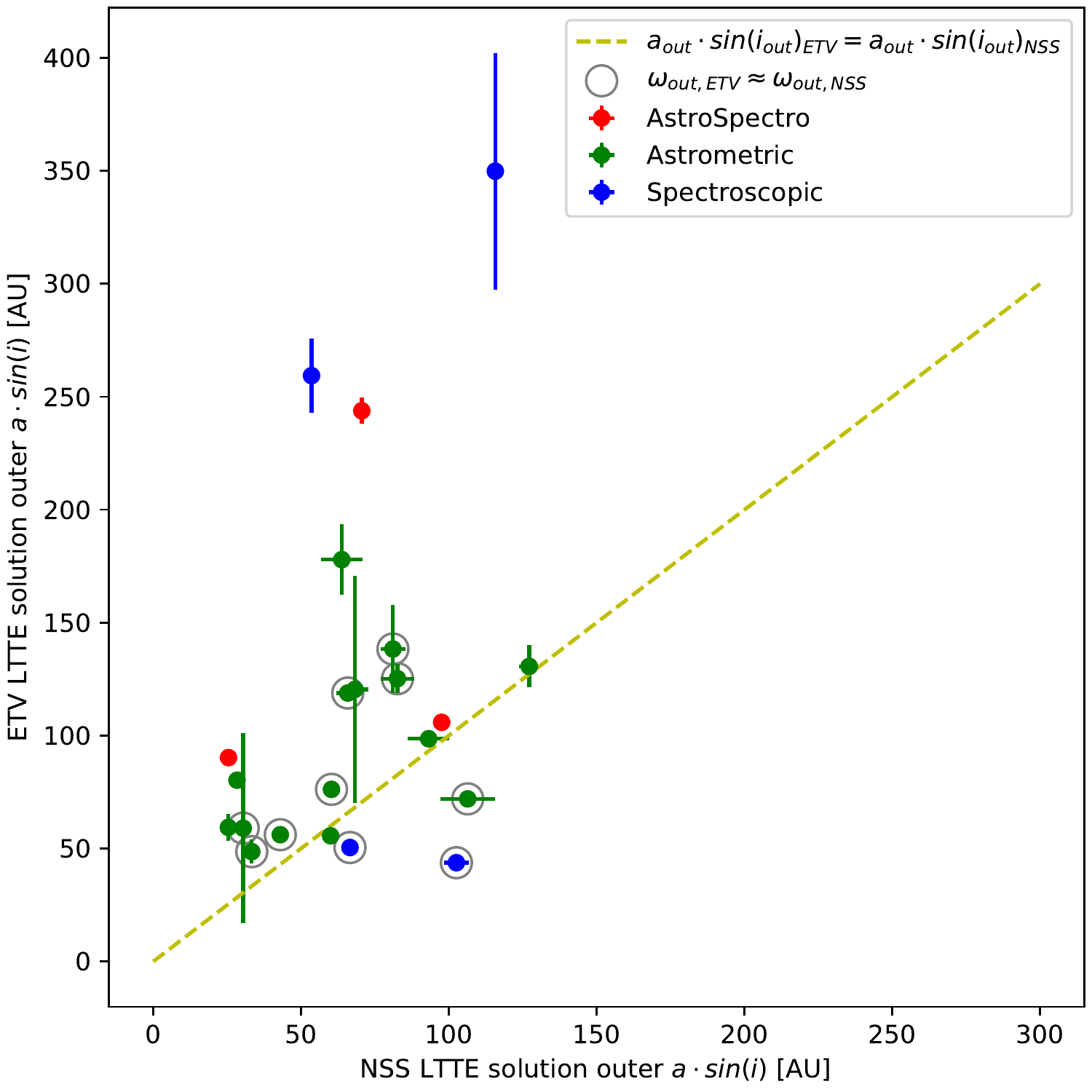}
    \includegraphics[scale = 0.6]{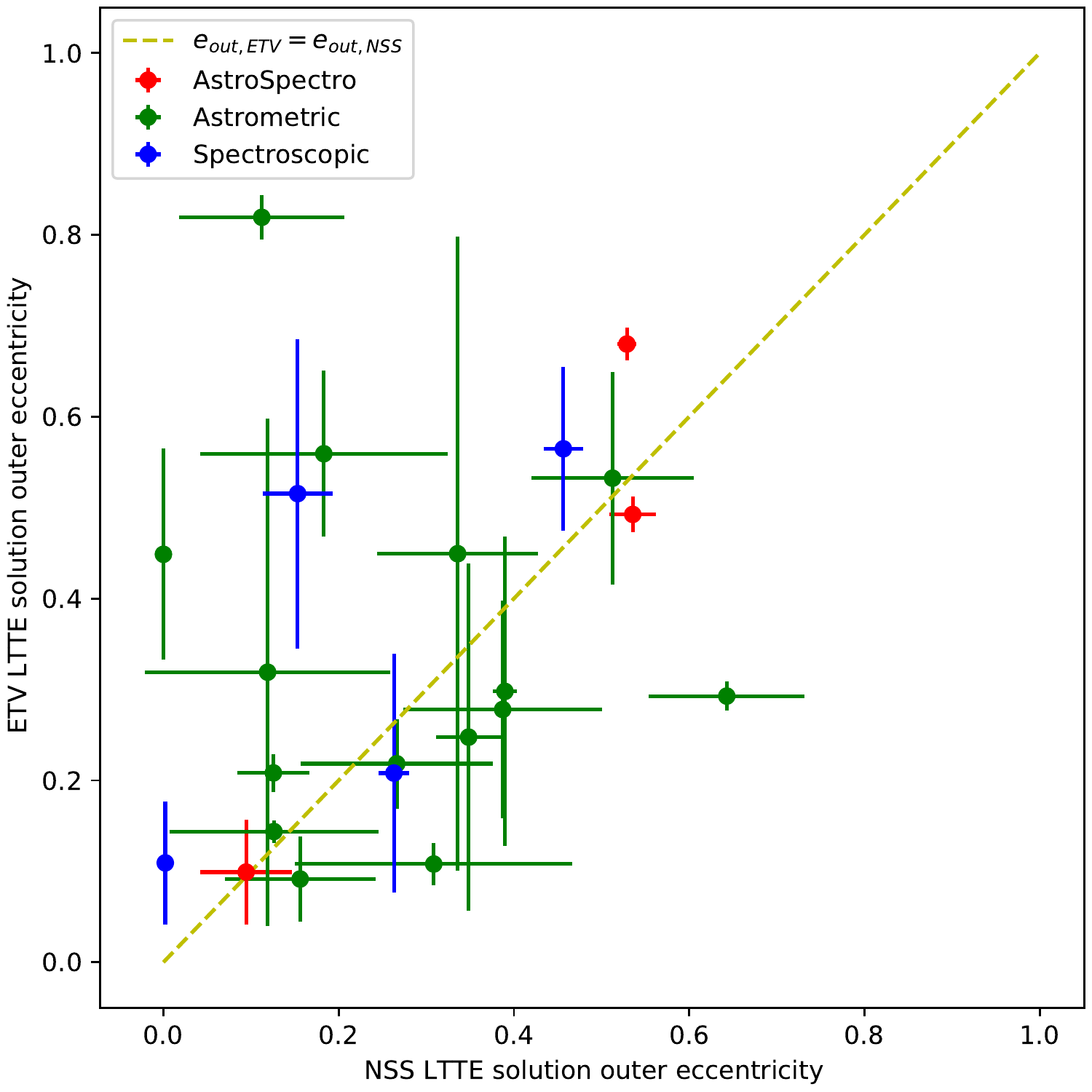}
    \includegraphics[scale = 0.6]{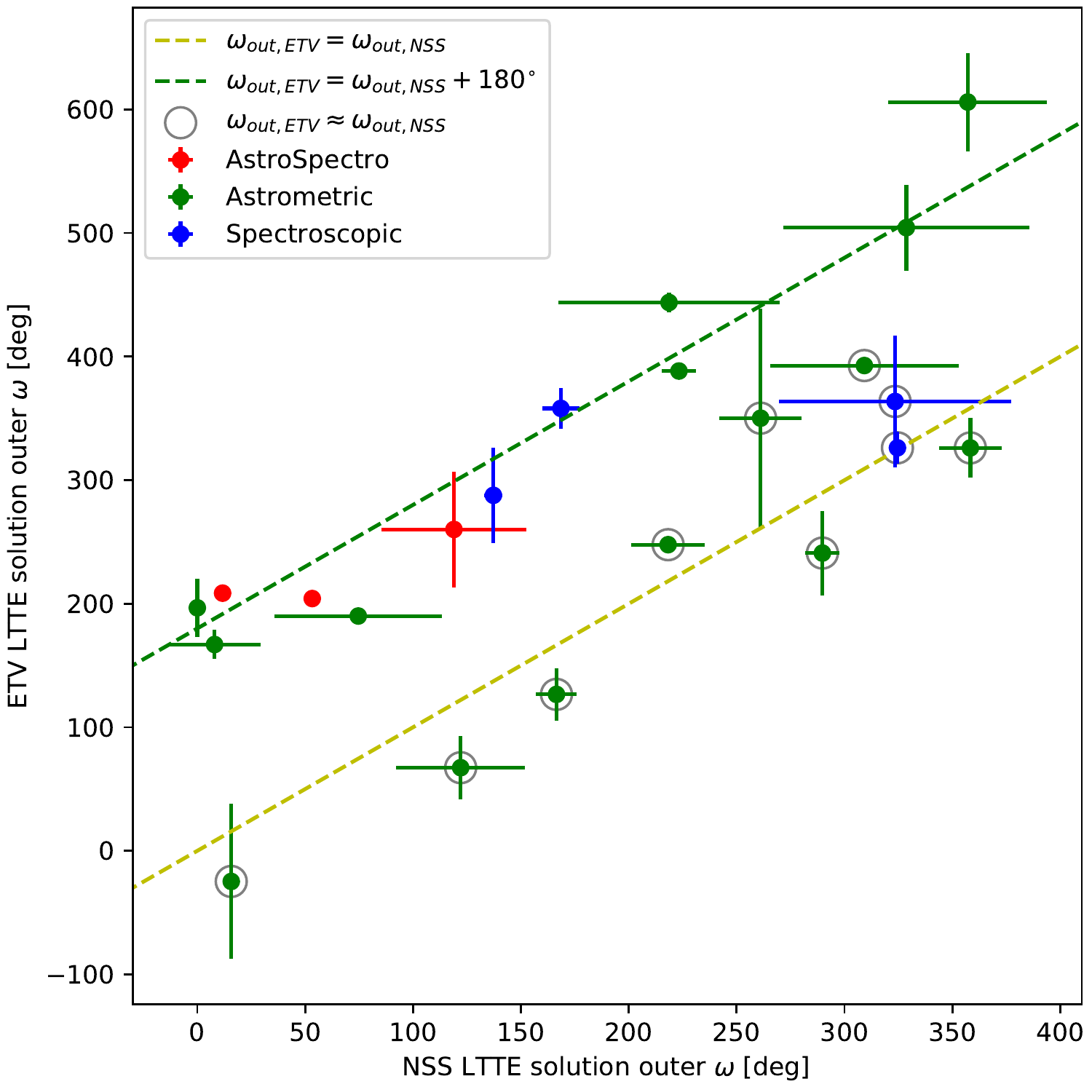}
\end{figure*}

\begin{figure*}
\caption{\textit{TESS} ETVs and their corresponding LTTE models for two of our triple candidates that have insufficient orbital coverage. Delay times calculated from primary and secondary eclipses are indicated with red and blue points, respectively, while their averaged values for the same orbital cycles are shown with black points. The green dashed lines show the LTTE models calculated with the orbital parameters from the \textit{Gaia} NSS solution, while the orange lines represent a sample of various LTTE models calculated with different orbital parameters that could fit the ETVs. $P_0$ and $T_0$ are the eclipsing period and the epoch values, respectively, used to calculate the ETVs.\vspace{0.2cm}}             
\label{unvalidated_etvs}
    \centering
    \includegraphics[scale = 0.6]{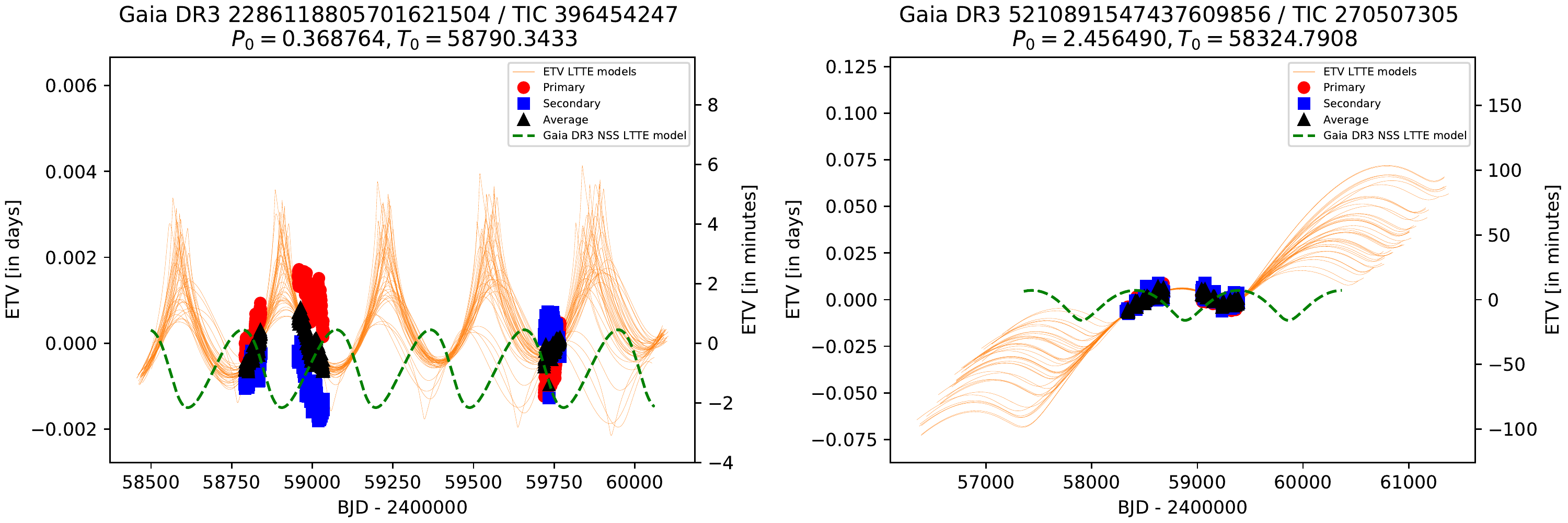}
\end{figure*}

As one can see in Fig.~\ref{validated_etvs_1} (and the other such figures in the Appendix), the most noticeable thing is that the LTTE models calculated with the parameters of the \textit{Gaia} orbital solutions can have either the same argument of periastron ($\omega_{NSS} = \omega_{ETV}$) or be 180 degrees away ($\omega_{NSS} = \omega_{ETV} \pm 180\degr$) from the LTTE models coming from our modeling process. This is not an error, and the reason for this is that the ETVs calculated from the \textit{TESS} eclipses are always tracking the motion of the EB around the center of mass of the triple system, while for \textit{Gaia}, this case is not so simple. For the astrometric solutions, \textit{Gaia} uses the variations in the position in the photocenter of the object, while for spectroscopic solutions, it uses the variations in one or two detected peaks (for SB1 and SB2 solutions, respectively) in the cross-correlation function (CCF). This means that without the ETV solutions (or in the case of spectroscopic solutions, the CCFs) it is not straightforward to say which component of the triple system is tracked by \textit{Gaia}.

In order to compare the orbital parameters of the \textit{Gaia} NSS and our own ETV solutions, we plot their correlations in Fig.~\ref{parameter_correlations}. On the top left panel, the two kinds of orbital periods are seen to be almost perfectly aligned with the line that represents a ratio of unity. That means \textit{Gaia} found the outer orbital periods of these triples almost perfectly. On the top right panel, the projected semi-major axis of the different solutions are compared and it can be seen that for this parameter, the relation is far from perfect. Only a handful of objects can be found near the line of unity, and almost all are more or less above this line. This could arise from two reasons related to the nature of the astrometric solutions: i) for the majority of triple system parameters, the photocenter revolves around a smaller orbit, hence the semi-major axis of the outer orbit will be underestimated \citep[see Eq. C2 of][]{rappaport22}; ii) the ETV solutions can overestimate the semi-major axis if the orbits are not completely covered. However, for spectroscopic solutions, we cannot really say anything certain without reviewing the individual spectra\footnote{These are not published in \textit{Gaia} DR3.} and the corresponding CCFs. In particular, the Gaia semi-major axis will depend on which component of the triple is being followed -- in which case the sign of the discrepancy could go either way. The bottom left panel shows the relation of the eccentricities from the different solutions. Most points are in an agreement within the uncertainties, however, for the majority of the solutions, the error bars are quite large. As there are a high number of ETV solutions in which only a limited part of the orbit is covered, in those cases the orbit can be fitted equally well with a relatively wide range of eccentricities that results in larger uncertainties. Finally, on the bottom right panel, the arguments of periastron of the two solutions are compared to each other and those are grouped along two lines as expected. Those systems in which \textit{Gaia} tracks the motion of the EB component (indicated with circles around them) are aligned with the yellow dashed line representing unity, while systems in which the tertiary component is tracked are aligned with the other, green dashed line. In conclusion, out of the 22 objects with the best-sampled \textit{TESS} ETVs, one can state that, for triple candidates, the most reliable parameter from the \textit{Gaia} orbital solutions is the outer orbital period and the outer eccentricity (although with higher uncertainties), while the other parameters should be used with caution, and only after identifying which component of the system is actually tracked by \textit{Gaia}.

We also found four additional systems (\object{TIC 66893949}, \object{TIC 298714297}, \object{TIC 88206187} and \object{TIC 14839347}) that show outer orbit third-body eclipses which inherently confirms the triple nature of these objects. First, we checked the FFIs of the corresponding objects and the extra eclipse events originate from the same pixels as the regular inner binary eclipses. Moreover, we also found evidence in archival ASAS-SN, ATLAS \citep{tonry18,smith20} and WASP \citep{butters10} photometric data that these objects are indeed triply eclipsing triple systems. The comprehensive photodynamical analysis of these newly discovered systems will be presented in a dedicated paper in the near future. Moreover, we also identified four additional systems (TIC 457671987, TIC 224053059, TIC 376606423 and TIC 410654298) that are most likely previously unknown EB+EB quadruples according to their \textit{TESS} LCs. Nevertheless, we do not consider them as validated systems because although the signal of both EBs for these objects comes from the same pixels that does not necessary confirm that they are certainly bound in the same system.  In particular, the two EBs could just be seen in the same direction without any direct physical connection between them.

We would like to highlight the special case of TIC 410654298 here which is the only object we kept although the ratio of the \textit{Gaia} and the EB period is a bit less than 5. This object is cataloged with a $\sim$20 day EB period and there are two separate types of \textit{Gaia} NSS solutions (astrometric and combined astrometric+spectroscopic) available for it with a similar period of $\sim$170 day. These resulted in a period ratio above 5 in the beginning. Nevertheless, after analyzing its \textit{TESS} LC, we found that its actual EB period is $\sim$36.4 day which yields a period ratio below our chosen limit. However, as there are two different types of \textit{Gaia} NSS solutions available for this systems, we would assume that the orbit found by \textit{Gaia} is valid. A plausible explanation for this object could be that it is a 2+2 system with an eclipsing and a non-eclipsing component, and \textit{Gaia} found the inner orbit of the latter, hence the 170 day period would not correspond to the outer orbit of the two binaries around the center of mass of the quadruple system.

\subsection{Candidates with significant ETVs, but insufficient orbital coverage}
\label{validated_insufficient}

There are 192 objects that show significant non-linear ETVs, but only a very short part of their outer orbit is covered by \textit{TESS} observations, hence they cannot be fitted reliably with a simple LTTE model. Although the LTTE models calculated with the parameters from the \textit{Gaia} orbital solutions do not perfectly match the observed ETVs, they are at least in accordance regarding the outer orbital periods for most of these systems. These systems also demonstrate that the most trustworthy parameter from the \textit{Gaia} NSS solutions is the orbital period. We plotted two systems in Fig.~\ref{unvalidated_etvs} as an example to illustrate how the duration of available observational data compares to the typical outer orbital solutions. As non-linear ETVs are apparent in these systems, we also consider these systems as confirmed hierarchical triple systems, nevertheless we could not validate their outer orbital parameters from the \textit{Gaia} NSS solutions.

\begin{figure*}
\caption{Distribution of inner binary periods and the corresponding outer orbital periods for the newly found triple candidates with SB1-type spectroscopic \textit{Gaia} NSS solutions. The dashed line shows the limiting period ratio of 5 chosen by us based on the requirement for minimal dynamical stability. The color scale represents the truth probability of the solutions derived using the empirical formula of \citep{bashi22}.}             
\label{spec_distribution}
    \centering
    \includegraphics[scale = 1.00]{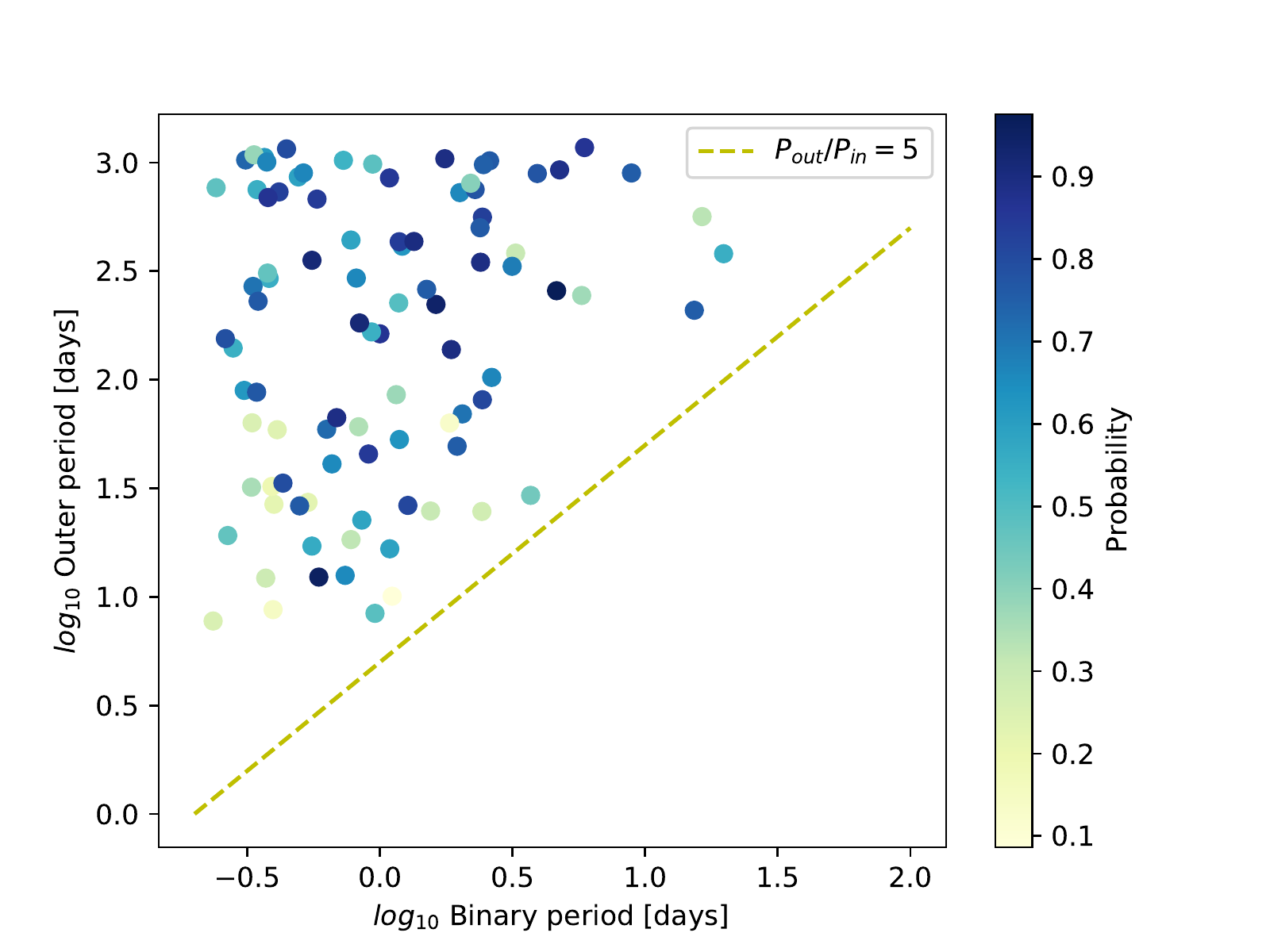}
\end{figure*}

\begin{table*}
\centering
\caption{Summary of the validation process for our newly identified hierarchical triple candidates.}
\begin{tabular}{lc}
\hline
\hline
Number & Category  \\
\hline
   403 & Total number of candidates  \\
   \hline
   27  & Previously known triples  \\
   22  &  Sufficient {\it TESS} ETV points to directly confirm outer orbit \\
   192 & Significant non-linear behavior in {\it TESS} ETV points confirms presence of third body \\
   4 & Exhibit third body eclipses, i.e., they are triply eclipsing triples \\
   4 & Probable quadruple system with 2+2 configuration \\
 \hline
  245 & Total confirmed triples \\
  217 & Total confirmed new triples \\
  \hline
  158 & Insufficient supplementary data to independently confirm triple nature \\
\hline 
\label{tbl:conffirm} 
\end{tabular}
\end{table*} 

\subsection{Candidates without significant ETVs or external data}
\label{unvalidated}
Finally, the remaining 158 systems (403 total minus 245 that are validated in one way or another) either do not have any or a sufficient amount of \textit{TESS} observations available, or their ETV curves do not show any non-linear behavior. In the absence of supplemental data, we could not confirm the triple nature of these systems, hence we consider them still as `candidates' that should be studied in the future when a sufficient amount of high-quality photometric (or spectroscopic) observations become available for them.

\subsection{Objects with short outer periods}
\label{short_period}
The shortest outer period CHT currently known is $\lambda$ Tau with a 33 day outer orbital period \citep{ebbighausen56}. As can be seen in our Fig.~\ref{period_distribution}, there are some 30 candidates with periods under 35 days. Empirically, such systems should be very rare, so it is somewhat surprising that a significant number of our systems could belong to this category. However, it is hard to confirm the true nature of these systems using only \textit{TESS} LCs and ETVs calculated from them. The amplitude of the LTTE depends on the semi-major axis which is obviously small for such systems, while the dynamical delays are proportional to the outer eccentricity which is also expected to be small according to formation scenarios, hence the amplitude of the ETVs of such systems are expected to be very small, within the order of our uncertainties. A further difficulty is that there is a very limited amount (or lack) of \textit{TESS} observations of these systems.

High-resolution spectroscopy with good temporal coverage could be of help in revealing the true configurations of these objects, however, that is not available for now as \textit{Gaia} DR3 contains only an average of all spectra for each object. Nevertheless, \citet{bashi22}, based on the spectroscopic observations of known EBs by LAMOST and GALAH, determined an empirical formula that can estimate the probability that a \textit{Gaia} spectroscopic SB1 solution is valid. We plot the period distribution of all 92 of our SB1-type candidates along with their empirical `truth probabilities' in Fig.~\ref{spec_distribution}. For 64 and 36 systems this probability is at least 50\% and 75\%, respectively. In general, it can be seen that the longer the outer period the more probable the solution is valid. This means that among the candidates with short outer periods, we expect there to be several invalid solutions. Also, there is a possibility that if the \textit{Gaia} orbital solution is valid, the object is actually a quadruple system with a 2+2 configuration with an eclipsing and a non-eclipsing binary component, and \textit{Gaia} actually found the spectroscopic orbital solution of the non-eclipsing binary component (see the case of SAO 167450 in Sect.~\ref{literature} and Table~\ref{known_candidates}). However, for now, we cannot confirm or reject these candidates with certainty, because of insufficient supplemental observational data.

\section{Conclusions}
\label{conclusions}
In this paper, we performed a search for possible hierarchical triple systems among more than a million cataloged EBs. We did this by looking for \textit{Gaia} astrometric and/or spectroscopic NSS solutions with periods that are at least 5 times longer than the corresponding EB period. We identified 403 such triple candidates of which 376 are newly proposed ones while 27 of them are previously known candidates confirmed by the \textit{Gaia} observations. We collected all available \textit{TESS} observations up to Sector 55 for all of our candidates and calculated their ETV curves. In the process, we discovered four newly found triply eclipsing triple systems simply from the extra eclipsing events found in their \textit{TESS} LCs and four additional 2+2 quadruple candidates. For 22 objects well-covered by \textit{TESS} observations, we could successfully fit the ETVs with a simple LTTE model using the parameters of the \textit{Gaia} orbital solutions after slight modifications. These model solutions showed that the most reliable parameter from the \textit{Gaia} solutions is the orbital period. However, for the projected semi-major axis, the eccentricity and the argument of periastron, we had to make larger changes in order to achieve acceptable fits to the ETVs, hence we think their values taken from the \textit{Gaia} solutions should be used with caution. There are also 192 objects among our candidates that show significant non-linear ETVs, however those could not be fitted reliably with a simple LTTE solution due to insufficient temporal coverage of the \textit{TESS} data. Nevertheless, we consider these systems as confirmed candidates as well because of their non-linear ETVs, even though they have no LTTE orbital solutions based on the ETVs. This makes a total of 218 reasonably confirmed CHTs which is more than half the number of the formerly known 201 listed systems in the latest version of the Multiple Star Catalog \citep{tokovinin18} and the 193 CHTs identified by \citet{hajdu19} and \citet{hajdu22}, collectively. The rest of the 158 systems, for now, remain in the status of triple candidates as there are no available supplemental \textit{TESS} observations for them or they do not show detectable non-linear ETVs. These systems should be revisited in the future after sufficient observational data become available for them in order to confirm their triple nature, along with those 192 systems for which the currently available ETV data do not allow us to determine their orbital parameters independently from the \textit{Gaia} orbital solutions. We summarize these numbers in Table~\ref{tbl:conffirm}

Finally, we would like to emphasize the efficiency of our newly applied method in finding previously unknown hierarchical triple or multiple systems. Out of the 1 million EBs that we utilized, we identified 403 triple or multiple star candidates (0.04 \%). This is eight times higher compared to the efficiency of the VSG \citep{vsg22} who visually inspected $10^6$ \textit{TESS} EB LCs to find 50 triply eclipsing triples ($\sim$0.005~\%). That means our newly introduced method is about an order of magnitude more productive in finding compact hierarchical triples, and it will be useful to identify further such objects when future \textit{Gaia} data releases and more cataloged EBs from ongoing and future large surveys will become available.

\begin{acknowledgements}
This research has made use of NASA’s Astrophysics Data System. This research has made use of the SIMBAD database, operated at CDS, Strasbourg, France.

This work has made use of data from the European Space Agency (ESA) mission
{\it Gaia} (\url{https://www.cosmos.esa.int/gaia}), processed by the {\it Gaia}
Data Processing and Analysis Consortium (DPAC,
\url{https://www.cosmos.esa.int/web/gaia/dpac/consortium}). Funding for the DPAC
has been provided by national institutions, in particular the institutions
participating in the {\it Gaia} Multilateral Agreement.

This paper includes data collected by the \textit{TESS} mission. Funding for the \textit{TESS} mission is provided by the NASA Science Mission directorate. Some of the data presented in this paper were obtained from the Mikulski Archive for Space Telescopes (MAST). STScI is operated by the Association of Universities for Research in Astronomy, Inc., under NASA contract NAS5-26555. Support for MAST for non-HST data is provided by the NASA Office of Space Science via grant NNX09AF08G and by other grants and contracts.

\end{acknowledgements}

%
%

\bibliographystyle{aa}
\bibliography{main}
\begin{appendix}

\section{Additional tables and figures}

\begin{sidewaystable*}[!h]
\caption{List of our newly identified triple system candidates and the parameters of their corresponding \textit{Gaia} NSS orbital solutions. The full table will be available in the on-line version of the paper.}             
\label{list_of_candidates}      
\centering  
\begin{tabular}{c c c c c c c c c c} \hline\hline 
Gaia DR3 source ID & \begin{tabular}{@{}c@{}} $\mathrm{RA}$ \\ $\mathrm{[^\circ]}$ \end{tabular} & \begin{tabular}{@{}c@{}} $\mathrm{DEC}$ \\ $\mathrm{[^\circ]}$ \end{tabular} & NSS model & \begin{tabular}{@{}c@{}} $\mathrm{P_{in}}$ \\ $\mathrm{[days]}$ \end{tabular} & \begin{tabular}{@{}c@{}} $\mathrm{P_{out}}$ \\ $\mathrm{[days]}$ \end{tabular} & \begin{tabular}{@{}c@{}} $\mathrm{a_{out}\sin{i_{out}}}$ \\ $\mathrm{[R_{\odot}]}$ \end{tabular} & \begin{tabular}{@{}c@{}} $\mathrm{e_{out}}$ \\  \end{tabular} & \begin{tabular}{@{}c@{}} $\mathrm{\omega_{out}}$ \\ $\mathrm{[^\circ]}$ \end{tabular}  & \begin{tabular}{@{}c@{}} $\mathrm{\tau_{out}}$ \\ $\mathrm{[days]}$ \end{tabular}
\\ \hline 
1020753820332905344 & 147.863 & 53.908 & SB1 & 0.395 & 8.733 (0.011) & 0.254 (0.036) & 0.060 (0.171) & 121.99 (196.93) & -3.53 (4.98)
\\ \hline 
1031356788916864640 & 124.364 & 51.863 & Orbital & 0.373 & 902.438 (12.343) & 57.950 (1.348) & 0.464 (0.020) & 105.67 (4.26) & -331.81 (3.66)
\\ \hline 
1066419871128922112 & 151.548 & 66.553 & Orbital & 1.637 & 408.272 (3.140) & 82.620 (5.755) & 0.513 (0.092) & 166.48 (9.41) & 47.34 (8.04)
\\ \hline 
1108427881102595840 & 109.812 & 69.054 & Orbital & 0.328 & 456.751 (5.395) & 31.094 (0.894) & 0.056 (0.050) & 141.02 (63.01) & 15.90 (76.33)
\\ \hline 
1120305973934059648 & 139.367 & 71.337 & Orbital & 0.326 & 791.852 (54.926) & 122.771 (11.144) & 0.046 (0.113) & 298.50 (178.17) & 82.63 (389.85)
\\ \hline 
1135425289609225984 & 115.974 & 74.688 & SB1 & 1.275 & 26.363 (0.004) & 8.858 (0.091) & 0.386 (0.010) & 225.03 (1.27) & 9.54 (0.07)
\\ \hline 
1137926506762647168 & 133.996 & 78.407 & Orbital & 0.248 & 449.180 (1.820) & 42.980 (3.068) & 0.348 (0.037) & 289.76 (8.01) & -21.51 (8.11)
\\ \hline 
1281813580534856448 & 222.532 & 29.650 & Orbital & 0.302 & 637.464 (8.543) & 39.734 (0.888) & 0.412 (0.033) & 148.89 (4.38) & -132.82 (6.02)
\\ \hline 
1325067886936436224 & 248.758 & 33.213 & Orbital & 0.421 & 454.296 (1.381) & 60.189 (0.760) & 0.339 (0.024) & 34.48 (3.46) & -85.00 (4.02)
\\ \hline 
1349068645222620672 & 266.297 & 43.605 & Orbital & 0.264 & 227.926 (0.873) & 77.006 (1.836) & 0.401 (0.062) & 194.02 (6.61) & -68.94 (4.44)
\\ \hline 
1349910664971031680 & 268.997 & 45.874 & Orbital & 0.967 & 550.137 (1.195) & 153.342 (1.142) & 0.332 (0.014) & 173.14 (2.63) & -221.92 (3.72)
\\ \hline 
1352346392464295168 & 251.723 & 38.650 & SB1 & 0.382 & 292.867 (3.469) & 97.300 (4.580) & 0.194 (0.051) & 228.37 (20.88) & -57.24 (15.26)
\\ \hline 
1374182281074944384 & 234.754 & 34.604 & Orbital & 0.545 & 507.810 (11.427) & 60.678 (4.277) & 0.162 (0.114) & 189.16 (38.86) & -74.69 (53.97)
\\ \hline 
1400153329838590080 & 240.980 & 48.954 & Orbital & 0.278 & 284.330 (3.579) & 25.420 (1.778) & 0.183 (0.141) & 218.65 (51.27) & -39.71 (42.79)
\\ \hline 
1428608033634193920 & 245.121 & 54.840 & Orbital & 4.128 & 241.031 (2.420) & 20.885 (1.561) & 0.084 (0.137) & 201.20 (89.06) & -37.79 (59.70)
\\ \hline 
1433502578364325120 & 255.971 & 57.365 & AstroSpectroSB1 & 0.511 & 607.032 (8.870) & 128.194 (3.081) & 0.631 (0.017) & 81.85 (1.33) & 19.51 (2.17)
\\ \hline 
1438455538945959680 & 258.253 & 61.623 & Orbital & 0.294 & 529.805 (6.714) & 93.175 (7.056) & 0.643 (0.089) & 223.26 (8.03) & -173.50 (7.73)
\\ \hline 
1445773647822084352 & 199.907 & 23.021 & AstroSpectroSB1 & 0.247 & 731.752 (10.597) & 51.417 (4.132) & 0.495 (0.029) & 346.24 (6.53) & -137.07 (5.07)
\\ \hline 
1458572925243347584 & 206.601 & 33.656 & Orbital & 0.651 & 263.061 (1.697) & 51.424 (2.047) & 0.346 (0.081) & 26.06 (10.35) & -21.93 (6.81)
\\ \hline 
1468504126582376448 & 202.295 & 31.402 & SB1 & 2.433 & 80.858 (0.050) & 53.545 (1.138) & 0.263 (0.017) & 137.25 (4.24) & 5.79 (0.84)
\\ \hline 
1487729293552139008 & 217.282 & 39.233 & Orbital & 0.357 & 611.335 (9.778) & 83.876 (9.598) & 0.554 (0.087) & 281.71 (7.58) & -73.23 (12.31)
\\ \hline 
1514368158189974784 & 189.724 & 31.950 & SB1 & 0.855 & 22.556 (0.006) & 10.397 (0.198) & 0.119 (0.020) & 109.91 (10.21) & 8.64 (0.60)
\\ \hline 
1526621047051675648 & 198.778 & 44.043 & Orbital & 3.473 & 255.545 (0.474) & 39.347 (1.761) & 0.529 (0.057) & 194.82 (4.27) & -52.22 (2.97)
\\ \hline 
1529088007547276416 & 196.372 & 42.931 & Orbital & 0.259 & 479.554 (2.538) & 74.556 (1.329) & 0.302 (0.026) & 188.40 (5.46) & -187.36 (7.01)
\\ \hline 
1548209614266676864 & 182.038 & 50.556 & SB1 & 0.312 & 1027.947 (107.855) & 217.753 (36.030) & 0.679 (0.106) & 126.97 (9.97) & -487.87 (14.15)
\\ \hline 
1575432933756377472 & 181.583 & 57.176 & Orbital & 0.416 & 2467.707 (553.037) & 68.097 (18.872) & 0.634 (0.059) & 44.13 (17.21) & -73.41 (5.68)
\\ \hline 
157816855305858176 & 71.459 & 28.661 & SB1 & 0.554 & 354.755 (0.944) & 119.834 (0.634) & 0.282 (0.009) & 99.24 (2.26) & 173.72 (2.91)
\\ \hline 
1617361920624734464 & 222.849 & 59.151 & Orbital & 1.014 & 170.163 (0.233) & 15.903 (3.505) & 0.837 (0.039) & 187.25 (8.72) & -65.01 (1.22)
\\ \hline 
1618523760817775616 & 221.015 & 61.098 & SB1 & 0.815 & 293.707 (0.622) & 45.917 (0.647) & 0.246 (0.017) & 25.39 (2.94) & 91.50 (2.19)
\\ \hline 
1621085176233505536 & 223.038 & 64.634 & AstroSpectroSB1 & 0.378 & 997.741 (36.858) & 40.274 (5.712) & 0.342 (0.033) & 353.74 (11.68) & 0.10 (14.44)
\\ \hline 
1651826696713169024 & 257.877 & 72.136 & Orbital & 2.100 & 523.310 (4.514) & 24.603 (1.431) & 0.159 (0.060) & 282.28 (18.80) & -256.96 (26.96)
\\ \hline 
1666172162359453824 & 202.471 & 65.405 & Orbital & 0.304 & 539.058 (10.208) & 63.795 (7.051) & 0.112 (0.094) & 74.60 (38.85) & -266.29 (62.40)
\\ \hline 
1703274357605001728 & 246.559 & 75.580 & Orbital & 0.351 & 527.445 (8.399) & 65.812 (3.865) & 0.126 (0.119) & 309.20 (43.79) & -201.56 (64.32)
\\ \hline 
170882562435866496 & 61.206 & 33.957 & SB2 & 0.373 & 10.470 (0.002) & 59.854 (4.743) & 0.158 (0.074) & 14.49 (12.45) & -3.71 (0.32)
\\ \hline 
1762108224080370816 & 310.340 & 13.661 & Orbital & 2.473 & 945.196 (158.613) & 142.851 (11.291) & 0.517 (0.164) & 173.67 (14.80) & 54.32 (26.31)
\\ \hline 
... & ... & ... & ... & ... & ... & ... & ... & ... & ...
\end{tabular}
\tablefoot{For the `NSS model' column we kept the original designations used by the \textit{Gaia} team.}
\end{sidewaystable*}

\newpage
\begin{table*}[!h]
\caption{Revised EB periods of some systems based on the \textit{TESS} light curves.}             
\label{revised_periods}      
\centering                          
\begin{tabular}{c | c | c | c | c}        
\hline\hline                 
Common name & \textit{Gaia} DR3 source ID & Catalog & \begin{tabular}{@{}c@{}} Catalog period \\ $\mathrm{[days]}$ \end{tabular} & \begin{tabular}{@{}c@{}} Revised period \\ $\mathrm{[days]}$ \end{tabular}\\    
\hline                        
   T-Dra0-00240 & 1433502578364325120 & VSX & 6.7255 &	0.510774 \\      
   TIC 71612483 &	1468504126582376448 & \textit{TESS} &	1.2168732 &	2.432884 \\
   ROTSE1 J142907.71+391359.6 &	1487729293552139008 & VSX	& 0.30255899 &	0.356598 \\
   CzeV3812	& 1831059117963533312 & VSX &	2.552908 &	5.099693 \\
   TIC 395459736 &	1980553185643943552 & \textit{TESS} &	3.6081353	& 10.645300 \\ 
   TIC 156091346 &	203211262307324160 & \textit{TESS} &	2.3154933 &	4.635822 \\
   KIC 2583777 &	2051670422357018880 & VSX &	0.479052 &	0.958116 \\
   KIC 3869326 &	2073175598494019584 & \textit{Kepler} &	8.586044 &	0.416574 \\
   NSVS 3410088 &	2179335267214938752 & VSX &	2.58401 &	5.165620 \\
   KELT KC03C02563 &	228021501672011264 & VSX &	0.58604341 &	1.172104 \\
   ASASSN-V J042708.29+453939.7 &	256744662398887808 & VSX &	2.954 &	5.908340 \\
   ASAS J072627-1613.8 &	3028470161558479232 & VSX &	0.356033 &	0.293739 \\
   TIC 44631957 &	320524407049379840 & \textit{TESS} &	15.3114276 &	7.649232 \\
   HD 47451 &	3355773092148236288 & VSX &	4.55025 &	9.100553 \\
   TIC 437230910 &	3549833939509628672 & \textit{TESS} &	0.380517 &	0.761006 \\
   NSVS 3900682 &	358411314297148416 & VSX &	4.43334098 &	2.216110 \\
   T-Cas0-13655 &	391068665069024512 & VSX &	1.845491 &	4.655776 \\
   T-Cas0-04329 &	392566990180396416 & VSX &	2.62607 & 	3.227946 \\
   TIC 393870193 &	4543911748032072832 & APASS, VSX &	0.167114 &	0.286254 \\
   TIC 267352521 &	4550017714059128576 & APASS, VSX &	0.7224 &	0.530320 \\
   TIC 394340319 &	4620501391559884672 & \textit{TESS} &	3.0375069 &	6.074511 \\
   TIC 394657952 &	5200940898488873472 & \textit{TESS} &	0.8723736 &	1.744773 \\
   TIC 270507305 &	5210891547437609856 & \textit{TESS} &	1.2282336 &	2.456490 \\
   TIC 410654298 & 5236430419447179776 & \textit{TESS} & 19.9520732 & 36.397144\\
   OGLE-GD-ECL-01161 &	5253655437337474432 & VSX &	0.38388394 &	3.925129 \\
   TIC 305536358 &	5739030825283977216 & \textit{TESS} &	0.5464738 &	1.091392 \\
   TIC 402918064 &	5795384949443391232 & \textit{TESS} &	1.2811614 &	2.562248 \\
   TIC 342782002 &	5832199626727378048 & \textit{TESS} &	1.0971318 &	2.199392 \\
   TIC 337370789 &	5857186956187793664 & \textit{TESS} &	0.8380236 &	0.449518 \\
   TIC 273571987 &	6377883399265729536 & \textit{TESS} &	0.7514777 &	1.502912 \\
   TIC 197760286 &	6575016247758836480 & \textit{TESS} &	0.6013042 &	1.202615 \\
   NSVS 4650785 &	978497178999963008 & VSX &	2.74598 &	1.372904 \\
\hline                                   
\end{tabular}
\end{table*}
\newpage

\begin{table*}[!h]
\caption{Outer orbital parameters from the \textit{Gaia} NSS solutions and after the simple LTTE fits of ETVs for our 22 validated triple candidates. For each object, the upper values belong to the former, while the lower values correspond to the latter solution, respectively.}             
\label{LTTE_parameters}      
\centering                          
\begin{tabular}{c | c | c | c | c | c | c} \hline\hline \begin{tabular}{@{}c@{}} Common name \\ \textit{Gaia} DR3 source ID \end{tabular} & $\mathrm{P_{out} / P_{in}}$ & \begin{tabular}{@{}c@{}} $\mathrm{P_{out}}$ \\ $\mathrm{[days]}$ \end{tabular} & \begin{tabular}{@{}c@{}} $\mathrm{a_{out}\sin{i_{out}}}$ \\ $\mathrm{[R_{\odot}]}$ \end{tabular} & \begin{tabular}{@{}c@{}} $\mathrm{e_{out}}$ \\  \end{tabular} & \begin{tabular}{@{}c@{}} $\mathrm{\omega_{out}}$ \\ $[^\circ]$ \end{tabular} & \begin{tabular}{@{}c@{}} $\mathrm{\tau_{out}}$ \\ $\mathrm{[days]}$ \end{tabular}\\ \hline \begin{tabular}{@{}c@{}} *\object{TIC 103606244} \\ 1066419871128922112 (A) \end{tabular} & 249.43 & \begin{tabular}{@{}c@{}} 408.27 (3.14) \\ 418.33 (16.11) \end{tabular} & \begin{tabular}{@{}c@{}} 82.62 (5.76) \\ 125.17 (6.48) \end{tabular} & \begin{tabular}{@{}c@{}} 0.51 (0.09) \\ 0.53 (0.12) \end{tabular} & \begin{tabular}{@{}c@{}} 166.48 (9.41) \\ 126.62 (21.18) \end{tabular} & \begin{tabular}{@{}c@{}} 47.34 (8.04) \\ -62.42 (90.55) \end{tabular}\\ \hline \begin{tabular}{@{}c@{}} *\object{V0500 Cam} \\ 1137926506762647168 (A) \end{tabular} & 1808.66 & \begin{tabular}{@{}c@{}} 449.18 (1.82) \\ 464.36 (12.82) \end{tabular} & \begin{tabular}{@{}c@{}} 42.98 (3.07) \\ 56.11 (2.04) \end{tabular} & \begin{tabular}{@{}c@{}} 0.35 (0.04) \\ 0.25 (0.19) \end{tabular} & \begin{tabular}{@{}c@{}} 289.76 (8.01) \\ 240.99 (34.14) \end{tabular} & \begin{tabular}{@{}c@{}} -21.51 (8.11) \\ -151.82 (51.46) \end{tabular}\\ \hline \begin{tabular}{@{}c@{}} \object{V1151 Her} \\ 1400153329838590080 (A) \end{tabular} & 1022.11 & \begin{tabular}{@{}c@{}} 284.33 (3.58) \\ 246.72 (1.15) \end{tabular} & \begin{tabular}{@{}c@{}} 25.42 (1.78) \\ 59.35 (5.92) \end{tabular} & \begin{tabular}{@{}c@{}} 0.18 (0.14) \\ 0.56 (0.09) \end{tabular} & \begin{tabular}{@{}c@{}} 218.65 (51.27) \\ 83.79 (7.89) \end{tabular} & \begin{tabular}{@{}c@{}} -39.71 (42.79) \\ 100.81 (7.88) \end{tabular}\\ \hline \begin{tabular}{@{}c@{}} \object{V0504 Dra} \\ 1438455538945959680 (A) \end{tabular} & 1800.17 & \begin{tabular}{@{}c@{}} 529.80 (6.71) \\ 543.98 (1.01) \end{tabular} & \begin{tabular}{@{}c@{}} 93.18 (7.06) \\ 98.60 (1.04) \end{tabular} & \begin{tabular}{@{}c@{}} 0.64 (0.09) \\ 0.29 (0.02) \end{tabular} & \begin{tabular}{@{}c@{}} 223.26 (8.03) \\ 28.32 (3.57) \end{tabular} & \begin{tabular}{@{}c@{}} -173.50 (7.73) \\ -254.05 (5.48) \end{tabular}\\ \hline \begin{tabular}{@{}c@{}} \object{TIC 71612483} \\ 1468504126582376448 (S) \end{tabular} & 33.24 & \begin{tabular}{@{}c@{}} 80.86 (0.05) \\ 79.49 (0.17) \end{tabular} & \begin{tabular}{@{}c@{}} 53.55 (1.14) \\ 259.33 (16.48) \end{tabular} & \begin{tabular}{@{}c@{}} 0.26 (0.02) \\ 0.21 (0.13) \end{tabular} & \begin{tabular}{@{}c@{}} 137.25 (4.24) \\ 287.73 (38.54) \end{tabular} & \begin{tabular}{@{}c@{}} 5.79 (0.84) \\ 70.32 (9.48) \end{tabular}\\ \hline \begin{tabular}{@{}c@{}} \object{TIC 158436971} \\ 1666172162359453824 (A) \end{tabular} & 1771.36 & \begin{tabular}{@{}c@{}} 539.06 (10.21) \\ 561.65 (1.50) \end{tabular} & \begin{tabular}{@{}c@{}} 63.80 (7.05) \\ 177.88 (15.70) \end{tabular} & \begin{tabular}{@{}c@{}} 0.11 (0.09) \\ 0.82 (0.02) \end{tabular} & \begin{tabular}{@{}c@{}} 74.60 (38.85) \\ 190.04 (3.08) \end{tabular} & \begin{tabular}{@{}c@{}} -266.29 (62.40) \\ -210.32 (5.49) \end{tabular}\\ \hline \begin{tabular}{@{}c@{}} *\object{TIC 320324245} \\ 1703274357605001728 (A) \end{tabular} & 1502.34 & \begin{tabular}{@{}c@{}} 527.45 (8.40) \\ 531.79 (1.04) \end{tabular} & \begin{tabular}{@{}c@{}} 65.81 (3.87) \\ 118.85 (0.76) \end{tabular} & \begin{tabular}{@{}c@{}} 0.13 (0.12) \\ 0.14 (0.01) \end{tabular} & \begin{tabular}{@{}c@{}} 309.20 (43.79) \\ 32.77 (6.16) \end{tabular} & \begin{tabular}{@{}c@{}} -201.56 (64.32) \\ -90.61 (7.76) \end{tabular}\\ \hline \begin{tabular}{@{}c@{}} *\object{V0499 Vul} \\ 1837166630181671808 (A) \end{tabular} & 298.38 & \begin{tabular}{@{}c@{}} 250.49 (2.54) \\ 246.84 (6.06) \end{tabular} & \begin{tabular}{@{}c@{}} 30.43 (2.44) \\ 59.01 (42.10) \end{tabular} & \begin{tabular}{@{}c@{}} 0.34 (0.09) \\ 0.45 (0.35) \end{tabular} & \begin{tabular}{@{}c@{}} 261.12 (19.06) \\ 350.16 (88.52) \end{tabular} & \begin{tabular}{@{}c@{}} -2.09 (14.60) \\ 81.95 (100.47) \end{tabular}\\ \hline \begin{tabular}{@{}c@{}} \object{TIC 28565554} \\ 2041591645936789504 (S) \end{tabular} & 661.98 & \begin{tabular}{@{}c@{}} 229.89 (1.33) \\ 226.38 (3.60) \end{tabular} & \begin{tabular}{@{}c@{}} 115.72 (2.71) \\ 349.84 (52.32) \end{tabular} & \begin{tabular}{@{}c@{}} 0.15 (0.04) \\ 0.52 (0.17) \end{tabular} & \begin{tabular}{@{}c@{}} 168.53 (8.48) \\ 358.09 (16.32) \end{tabular} & \begin{tabular}{@{}c@{}} -96.82 (5.49) \\ -151.66 (27.62) \end{tabular}\\ \hline \begin{tabular}{@{}c@{}} \object{TIC 170558954} \\ 2071891643809106560 (A) \end{tabular} & 1332.57 & \begin{tabular}{@{}c@{}} 495.19 (3.83) \\ 510.84 (15.96) \end{tabular} & \begin{tabular}{@{}c@{}} 127.19 (3.56) \\ 130.63 (9.34) \end{tabular} & \begin{tabular}{@{}c@{}} 0.00 (0.00) \\ 0.45 (0.12) \end{tabular} & \begin{tabular}{@{}c@{}} 0.00 (1.22) \\ 196.68 (23.68) \end{tabular} & \begin{tabular}{@{}c@{}} -101.99 (5.40) \\ -629.03 (146.75) \end{tabular}\\ \hline \begin{tabular}{@{}c@{}} *\object{TIC 377105433} \\ 2155105088944426624 (A) \end{tabular} & 2777.60 & \begin{tabular}{@{}c@{}} 698.29 (20.98) \\ 731.01 (2.05) \end{tabular} & \begin{tabular}{@{}c@{}} 106.38 (9.30) \\ 71.97 (2.88) \end{tabular} & \begin{tabular}{@{}c@{}} 0.31 (0.16) \\ 0.11 (0.02) \end{tabular} & \begin{tabular}{@{}c@{}} 122.09 (29.91) \\ 67.29 (25.81) \end{tabular} & \begin{tabular}{@{}c@{}} 260.89 (59.20) \\ 649.78 (48.60) \end{tabular}\\ \hline \begin{tabular}{@{}c@{}} *\object{TIC 331549024} \\ 2223170012666814720 (S) \end{tabular} & 500.44 & \begin{tabular}{@{}c@{}} 139.94 (0.09) \\ 143.10 (8.76) \end{tabular} & \begin{tabular}{@{}c@{}} 102.56 (4.28) \\ 43.72 (3.54) \end{tabular} & \begin{tabular}{@{}c@{}} 0.46 (0.02) \\ 0.56 (0.09) \end{tabular} & \begin{tabular}{@{}c@{}} 324.54 (2.94) \\ 326.16 (12.92) \end{tabular} & \begin{tabular}{@{}c@{}} -18.03 (0.65) \\ -47.27 (86.76) \end{tabular}\\ \hline \begin{tabular}{@{}c@{}} \object{TIC 293951386} \\ 2235168051048793600 (AS) \end{tabular} & 1220.42 & \begin{tabular}{@{}c@{}} 326.36 (2.33) \\ 344.30 (1.51) \end{tabular} & \begin{tabular}{@{}c@{}} 25.47 (1.58) \\ 90.24 (2.39) \end{tabular} & \begin{tabular}{@{}c@{}} 0.09 (0.05) \\ 0.10 (0.06) \end{tabular} & \begin{tabular}{@{}c@{}} 118.97 (33.71) \\ 260.11 (46.90) \end{tabular} & \begin{tabular}{@{}c@{}} -115.54 (30.85) \\ -183.48 (55.21) \end{tabular}\\ \hline \begin{tabular}{@{}c@{}} *\object{TIC 229762991} \\ 2259293058444658432 (A) \end{tabular} & 1257.82 & \begin{tabular}{@{}c@{}} 425.72 (2.63) \\ 424.32 (0.98) \end{tabular} & \begin{tabular}{@{}c@{}} 60.30 (2.74) \\ 76.19 (0.75) \end{tabular} & \begin{tabular}{@{}c@{}} 0.13 (0.04) \\ 0.21 (0.02) \end{tabular} & \begin{tabular}{@{}c@{}} 218.24 (17.06) \\ 247.78 (6.38) \end{tabular} & \begin{tabular}{@{}c@{}} -160.11 (19.93) \\ -141.83 (9.89) \end{tabular}\\ \hline \begin{tabular}{@{}c@{}} \object{TIC 229751802} \\ 2259381809648968320 (AS) \end{tabular} & 421.41 & \begin{tabular}{@{}c@{}} 217.56 (0.36) \\ 217.77 (0.18) \end{tabular} & \begin{tabular}{@{}c@{}} 97.58 (1.77) \\ 105.86 (1.69) \end{tabular} & \begin{tabular}{@{}c@{}} 0.54 (0.03) \\ 0.49 (0.02) \end{tabular} & \begin{tabular}{@{}c@{}} 11.73 (3.03) \\ 208.55 (2.15) \end{tabular} & \begin{tabular}{@{}c@{}} -99.02 (1.46) \\ -96.07 (2.08) \end{tabular}\\ \hline \begin{tabular}{@{}c@{}} \object{TIC 237234024} \\ 2288385277122496896 (A) \end{tabular} & 90.13 & \begin{tabular}{@{}c@{}} 170.88 (0.94) \\ 172.60 (0.34) \end{tabular} & \begin{tabular}{@{}c@{}} 28.35 (1.36) \\ 80.24 (2.05) \end{tabular} & \begin{tabular}{@{}c@{}} 0.27 (0.11) \\ 0.22 (0.05) \end{tabular} & \begin{tabular}{@{}c@{}} 8.01 (21.21) \\ 166.98 (11.82) \end{tabular} & \begin{tabular}{@{}c@{}} 65.69 (10.24) \\ 50.90 (6.31) \end{tabular}\\ \hline \begin{tabular}{@{}c@{}} \object{TIC 398357038} \\ 2303960099848580608 (A) \end{tabular} & 461.28 & \begin{tabular}{@{}c@{}} 676.95 (22.91) \\ 619.40 (7.37) \end{tabular} & \begin{tabular}{@{}c@{}} 68.20 (4.54) \\ 120.41 (50.16) \end{tabular} & \begin{tabular}{@{}c@{}} 0.12 (0.14) \\ 0.32 (0.28) \end{tabular} & \begin{tabular}{@{}c@{}} 328.66 (57.06) \\ 144.32 (34.63) \end{tabular} & \begin{tabular}{@{}c@{}} 14.55 (105.31) \\ -573.78 (90.34) \end{tabular}\\ \hline \begin{tabular}{@{}c@{}} *\object{TIC 269762258} \\ 4622145951717126656 (A) \end{tabular} & 328.30 & \begin{tabular}{@{}c@{}} 434.86 (0.72) \\ 434.56 (7.33) \end{tabular} & \begin{tabular}{@{}c@{}} 33.34 (1.51) \\ 48.60 (5.31) \end{tabular} & \begin{tabular}{@{}c@{}} 0.39 (0.01) \\ 0.30 (0.17) \end{tabular} & \begin{tabular}{@{}c@{}} 15.74 (3.29) \\ 335.22 (62.80) \end{tabular} & \begin{tabular}{@{}c@{}} -8.46 (2.43) \\ -60.33 (58.30) \end{tabular}\\ \hline \begin{tabular}{@{}c@{}} \object{TIC 373844472} \\ 4660192386982447232 (AS) \end{tabular} & 95.30 & \begin{tabular}{@{}c@{}} 175.26 (0.12) \\ 175.56 (0.10) \end{tabular} & \begin{tabular}{@{}c@{}} 70.56 (0.53) \\ 243.77 (5.78) \end{tabular} & \begin{tabular}{@{}c@{}} 0.53 (0.01) \\ 0.68 (0.02) \end{tabular} & \begin{tabular}{@{}c@{}} 53.28 (0.91) \\ 204.18 (1.20) \end{tabular} & \begin{tabular}{@{}c@{}} 16.78 (0.46) \\ 21.18 (1.03) \end{tabular}\\ \hline \begin{tabular}{@{}c@{}} \object{TIC 350336454} \\ 4766586320357214848 (A) \end{tabular} & 405.78 & \begin{tabular}{@{}c@{}} 469.48 (8.24) \\ 474.20 (2.56) \end{tabular} & \begin{tabular}{@{}c@{}} 60.00 (2.76) \\ 55.64 (1.62) \end{tabular} & \begin{tabular}{@{}c@{}} 0.16 (0.09) \\ 0.09 (0.05) \end{tabular} & \begin{tabular}{@{}c@{}} 357.13 (36.76) \\ 246.02 (39.77) \end{tabular} & \begin{tabular}{@{}c@{}} -201.69 (48.41) \\ -128.27 (55.60) \end{tabular}\\ \hline \begin{tabular}{@{}c@{}} *\object{TIC 287531033} \\ 5210208510199765120 (A) \end{tabular} & 787.43 & \begin{tabular}{@{}c@{}} 663.27 (14.83) \\ 661.07 (4.96) \end{tabular} & \begin{tabular}{@{}c@{}} 81.09 (4.24) \\ 138.32 (19.42) \end{tabular} & \begin{tabular}{@{}c@{}} 0.39 (0.11) \\ 0.28 (0.12) \end{tabular} & \begin{tabular}{@{}c@{}} 358.33 (14.60) \\ 326.10 (24.29) \end{tabular} & \begin{tabular}{@{}c@{}} 4.18 (24.63) \\ -65.74 (40.73) \end{tabular}\\ \hline \begin{tabular}{@{}c@{}} *\object{TIC 231922417} \\ 5566728797637156352 (S) \end{tabular} & 97.27 & \begin{tabular}{@{}c@{}} 66.85 (0.01) \\ 66.90 (0.08) \end{tabular} & \begin{tabular}{@{}c@{}} 66.56 (0.14) \\ 50.44 (1.87) \end{tabular} & \begin{tabular}{@{}c@{}} 0.00 (0.00) \\ 0.11 (0.07) \end{tabular} & \begin{tabular}{@{}c@{}} 323.44 (53.74) \\ 3.70 (53.45) \end{tabular} & \begin{tabular}{@{}c@{}} 27.64 (9.98) \\ 33.52 (10.43) \end{tabular}\end{tabular}
\tablefoot{Those systems in which \textit{Gaia} tracks the same component as the ETVs are indicated with an asterisk before their names. The \textit{Gaia} NSS solution type of the objects are noted after their DR3 source IDs (A: astrometric solution, S: spectroscopic solution, AS: combined solution). $\mathrm{\tau_{\mathrm{out}}}$ is the epoch of periastron passage relative to the \textit{Gaia} reference epoch of J2016.0.}
\end{table*}

\newpage

\begin{figure*}[!h]
\caption{\textit{TESS} ETVs and their corresponding LTTE models for eight of our validated triple candidates. Delay times calculated from primary and secondary eclipses are indicated with red and blue points, respectively, while their averaged values for the same orbital cycles are shown with black points. The green dashed lines denote the LTTE models calculated with the orbital parameters from the \textit{Gaia} NSS solution, while the blue lines represent our LTTE models after modifications in the orbital parameters. $P_0$ and $T_0$ are the eclipsing period and the epoch values, respectively, used to calculate the ETVs.\vspace{0.2cm}}             
\label{validated_etvs_2}
    \centering
    \includegraphics[scale = 0.48]{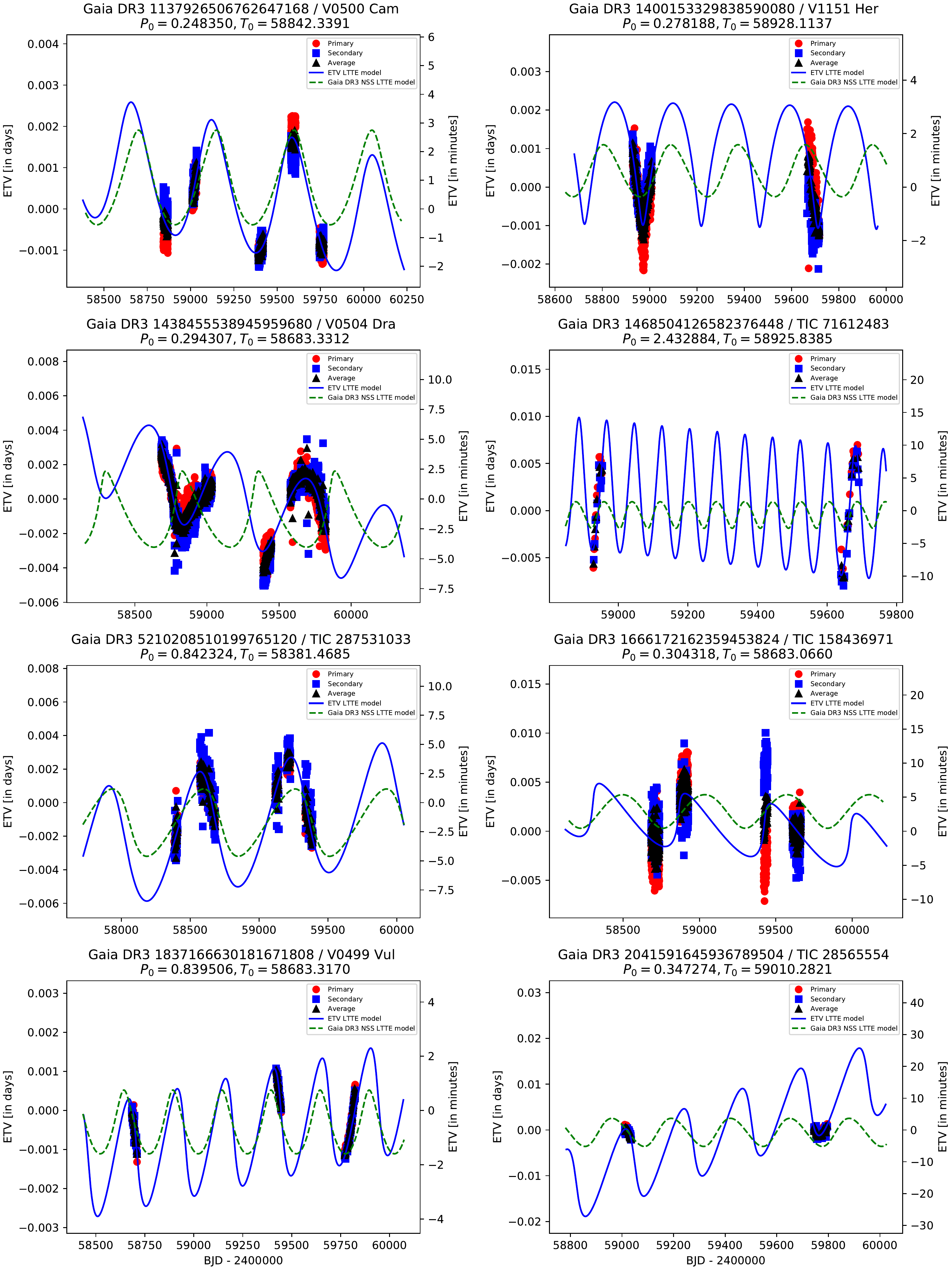}
\end{figure*}

\begin{figure*}[!h]
\caption{\textit{TESS} ETVs and their corresponding LTTE models for an additional eight systems among our validated triple candidates. Delay times calculated from primary and secondary eclipses are indicated with red and blue points, respectively, while their averaged values for the same orbital cycles are shown with black points. The green dashed lines denote the LTTE models calculated with the orbital parameters from the \textit{Gaia} NSS solution, while the blue lines represent our LTTE models after modifications in the orbital parameters. $P_0$ and $T_0$ are the eclipsing period and the epoch values, respectively, used to calculate the ETVs.\vspace{0.2cm}}             
\label{validated_etvs_3}
    \centering
    \includegraphics[scale = 0.48]{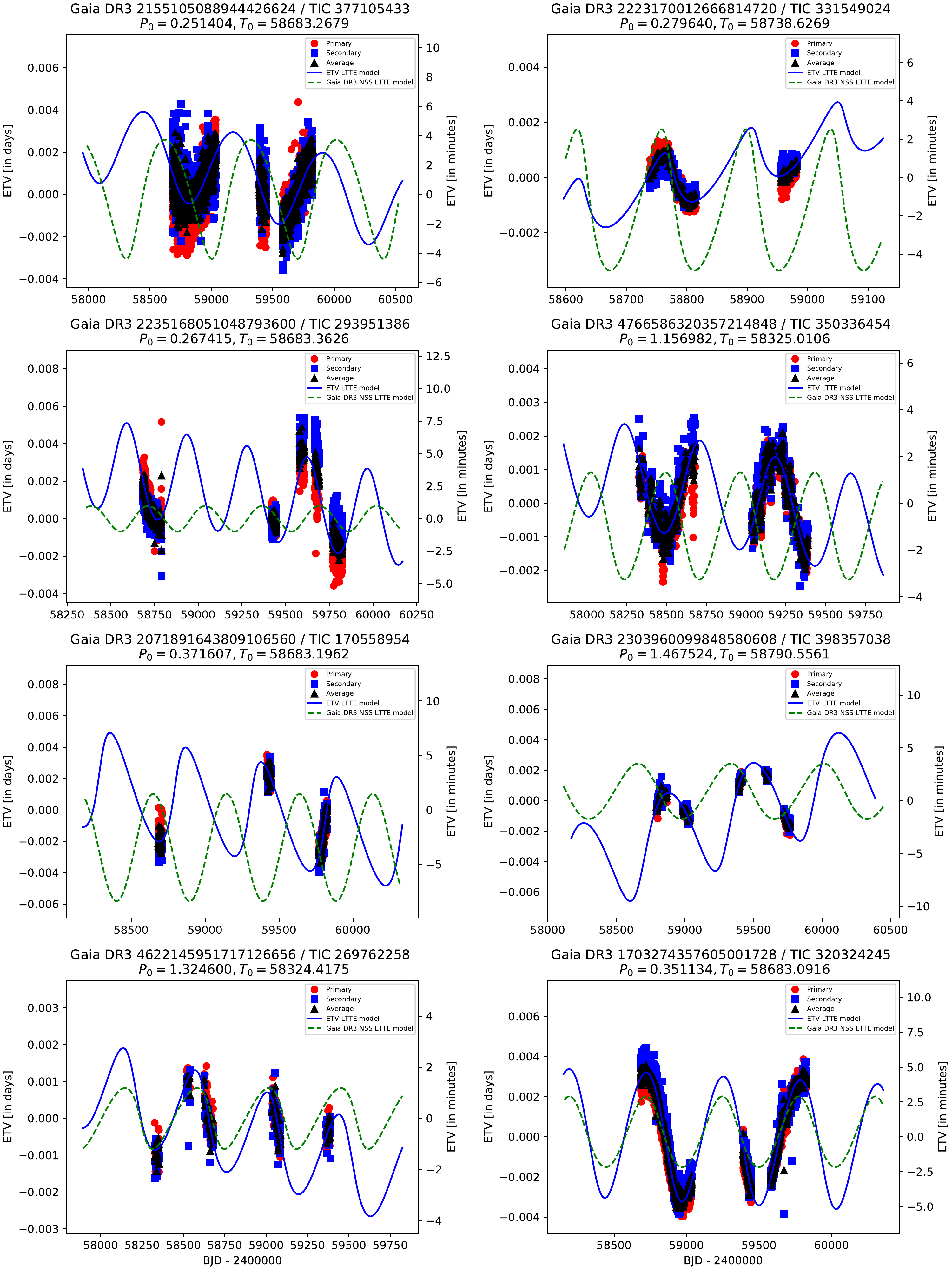}
\end{figure*}

\begin{figure*}[!h]
\caption{\textit{TESS} ETVs and their corresponding LTTE models for an additional four systems among our validated triple candidates. Delay times calculated from primary and secondary eclipses are indicated with red and blue points, respectively, while their averaged values for the same orbital cycles are shown with black points. The green dashed lines denote the LTTE models calculated with the orbital parameters from the \textit{Gaia} NSS solution, while the blue lines represent our LTTE models after modifications in the orbital parameters. $P_0$ and $T_0$ are the eclipsing period and the epoch values, respectively, used to calculate the ETVs.\vspace{0.2cm}}             
\label{validated_etvs_4}
    \centering
    \includegraphics[scale = 0.55]{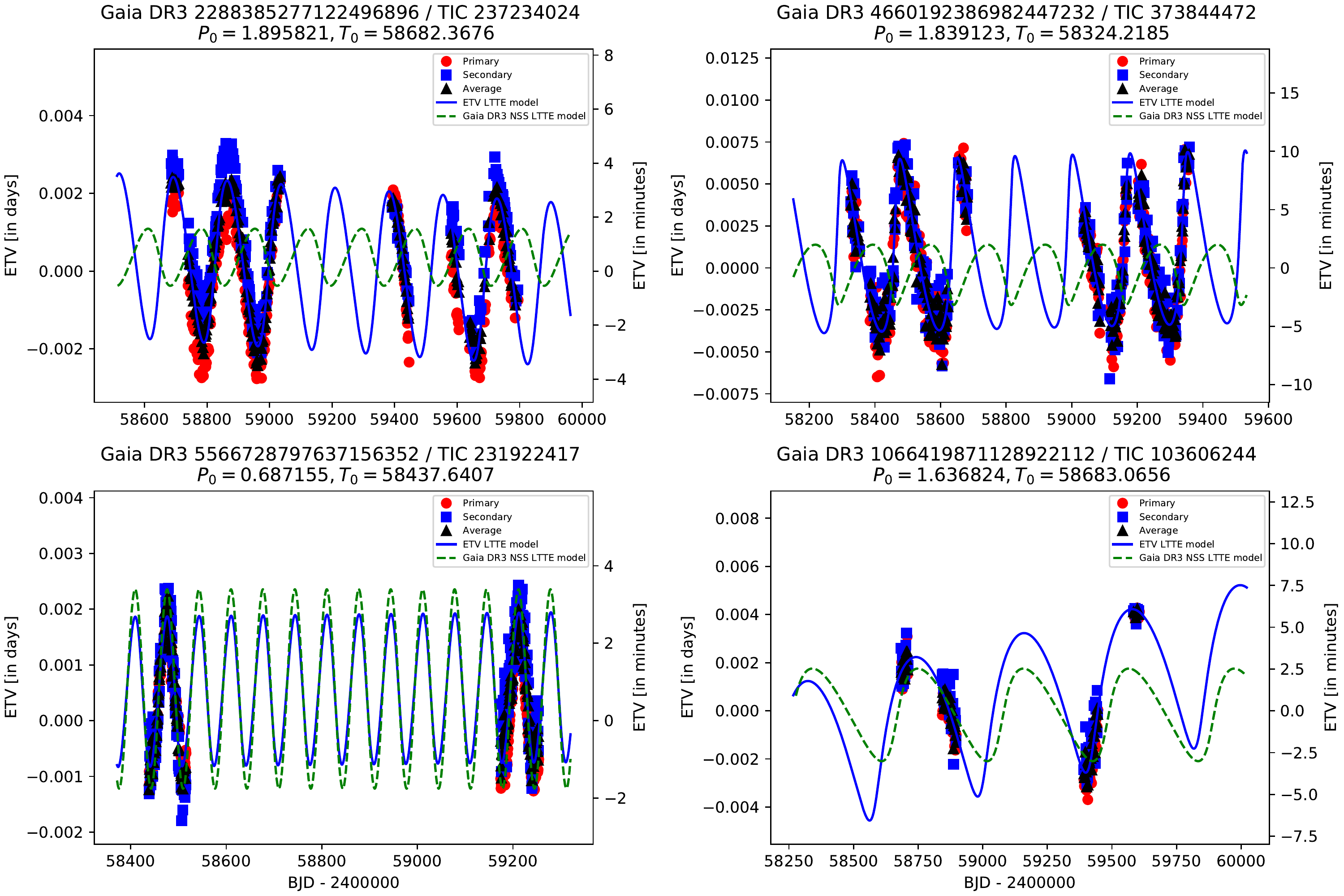}
\end{figure*}

\end{appendix}

\end{document}